\documentclass{aa} 

\usepackage{graphicx}
\usepackage{txfonts}
\usepackage{ulem}
\usepackage{multirow}
\usepackage{color}
\usepackage{times}
\usepackage{natbib}
\bibpunct{(}{)}{;}{a}{}{,} 
\usepackage{setspace}
\usepackage{changepage}
\usepackage{multicol}
\usepackage{latexsym}
\usepackage{float}
\usepackage[caption = false]{subfig}

\begin{document} 

\title{Bright submillimeter galaxies do trace galaxy protoclusters}



   \author{Rosa Calvi
          \inst{1,2,3,5}\fnmsep\thanks{\email{rosa.calvi@unife.it} (RC);\\ \texttt{gianluca.castignani@unibo.it} (GC); \texttt{helmut@iac.es} (HD)}
          \and
          Gianluca Castignani\inst{4,5}
          \and
          Helmut Dannerbauer\inst{1,2}
          }

   \institute{Instituto de Astrof\'isica de Canarias, E-38205 La Laguna, Spain  
   \and
            Depto. de Astrof\'isica, Universidad de La Laguna, E-38206 La Laguna, Spain
\and
Dipartimento di Fisica e Scienze della Terra, Università degli Studi di Ferrara, Via Saragat 1, I-44122 Ferrara, Italy 
\and
   Dipartimento di Fisica e Astronomia ''Augusto Righi'', Alma Mater Studiorum Università di Bologna, Via Gobetti 93/2, I-40129 Bologna, Italy   
            \and
            INAF - Osservatorio  di  Astrofisica  e  Scienza  dello  Spazio  di  Bologna,  via  Gobetti  93/3,  I-40129,  Bologna,  Italy            
             }

   \date{Accepted June 5, 2023}

 
  \abstract{There is controversy in the literature regarding whether distant, massive, and dusty starbursts selected at (sub)millimeter wavelengths can trace galaxy overdensities. We thus performed the first systematic search for distant protoclusters around a homogeneously selected sample of 12 spectroscopically confirmed submillimeter galaxies (SMGs) at $z\sim1.2-5.3$, which we selected from the GOODS-N field. We applied the well-established Poisson probability method (PPM) to search for megaparsec-scale overdensities around {these} SMGs, using three different photometric redshift catalogs. We robustly detect galaxy overdensities for 11 out of the 12 SMGs (i.e., $92\%\pm8$\%), distributed over eight large-scale protoclusters.  We confirm all three previously discovered protoclusters, and we detect five new ones around the SMGs SMM~J123634 ($z=1.225$), ID.19 ($z=2.047$), SMM~J123607 ($z=2.487$), SMM~J123606 ($z=2.505$), and GN10 ($z=5.303$). A wavelet-based analysis of the protocluster fields shows that the SMGs are located in protocluster cores with a complex morphology (compact, filamentary, or clumpy) and an average size of $\sim(0.4-1)$~Mpc. By comparing the PPM results obtained using the three redshift catalogs independently, each of which trace different galaxy populations and redshift ranges, we speculate that we are possibly witnessing a transitioning phase at $z\gtrsim4$  for the galaxy population of protoclusters. While $z\lesssim4$ protoclusters appear to be populated by dusty galaxies, those at the highest redshifts, {$z\sim5$}, are detected as overdensities of Lyman~$\alpha$ emitters or Lyman break galaxies. Further investigation with larger samples is required to reach a definitive conclusion. We also find a good correlation between the molecular (H$_2$) gas mass of the SMGs and the significance of the associated overdensity. To explain the overall phenomenology, we suggest that galaxy interactions in dense environments likely triggered the starburst and gas-rich phase of the SMGs. Altogether, our findings support the scenario that SMGs are excellent tracers of distant protoclusters. The ones presented in this work are excellent targets for the {\it James Webb} Space Telescope. Similarly, future surveys with forthcoming facilities (e.g., {\it Euclid} and LSST) can be tuned to detect even larger samples of distant protoclusters.}

    \keywords{Galaxies: clusters: general; Galaxies: evolution; Galaxies: high-redshift; Galaxies: starbursts; Infrared: galaxies; submillimeter: galaxies}

   \authorrunning{Calvi, Castignani \& Dannerbauer}
   \titlerunning{Bright SMGs do trace galaxy protoclusters}
   \maketitle
%
\section{Introduction}
Understanding when and how present-day galaxy clusters formed at high redshifts has been the main science driver for the extensive search for protoclusters of galaxies in the distant universe, especially at optical and near-infrared wavelengths (see \citealt{Overzier16} and \citealt{Alberts22} for reviews; the latter focuses on the infrared). The most massive galaxies in the present-day universe lie in rich clusters. Their old, coeval, metal-rich populations suggest that they may have formed as spectacular starbursts at high redshifts. Powerful high-redshift radio galaxies \citep[see the review by][]{Miley08} are considered to be the most promising signposts of massive clusters in formation.

\begin{table*}[ht!]
\small\addtolength{\tabcolsep}{-5pt}
\centering
Properties of the GOODS-N SMGs.\\
        \begin{tabular}{c|cccccccccc}
    \hline
    \hline
Name & $z_{\rm spec}$ & CO(J$\rightarrow$J-1) & ${\rm S}_{{\rm CO(J}\rightarrow {\rm J}-1)} \Delta\varv$ & $L^\prime_{\rm CO(J\rightarrow J-1)}$ & $\log(M_\star/M_\odot)$ & $M_{\rm H_2}$ & $\log(L_{\rm FIR}/L_\odot)$ & SFR & $\tau_{\rm dep}$& Reference \\
& & & (Jy~km~s$^{-1}$) & ($10^{10}$~K~km~s$^{-1}$~pc$^{2}$) & & ($10^{10}~M_\sun$) &  & (M$_{\odot}$~$yr^{-1}$)& (yr) &\\
(1) & (2) & (3) & (4) & (5) & (6) & (7) & (8) & (9) & (10) & (11) \\
\hline
\hline
SMM~J123634 & 1.225 & CO(2$\rightarrow$1) & $1.5\pm0.2$ & 2.97$\pm$0.40 & 10.82$^{a}$ & 2.64$\pm$0.35 & 12.56 & 330&  $8.0\times10^7$  &\citet{Bothwell13} \\
ID.03 &1.784 &   CO(2$\rightarrow$1)      &   $0.50\pm0.08$ & 2.00$\pm$0.32 & --- &  1.78$\pm$0.28 & 11.83  & 61& $2.9\times10^8$
  & \citet{Decarli14}\\ 
SMM~J123711 & 1.995 &  CO(3$\rightarrow$2)  &  $1.9\pm0.5$ & 4.14$\pm$1.09 & --- & 5.52$\pm$1.45 & 12.82 &601 &   $9.2\times10^7$  &\citet{Bothwell13}\\
SMM~J123618 & 1.996 & CO(4$\rightarrow$3) & $1.5\pm0.2$ & 1.84$\pm$0.25 & 10.71$^{a}$ & 4.60$\pm$0.61 & 12.89 & 706 &  $6.5\times10^7$  &\citet{Bothwell13} \\
SMM~J123712 & 1.996 & CO(3$\rightarrow$2)   &    $1.2\pm0.4$ & 2.62$\pm$0.87 & 10.47$^{a}$ & 3.49$\pm$1.16 & 12.43 &245 &  $1.4\times10^8$ &\citet{Bothwell13} \\
ID.19 & 2.047 &  CO(3$\rightarrow$2) &       $0.43\pm0.11$ & 0.98$\pm$0.25 & --- & 1.31$\pm$0.33 & 10.90  & 7.22&  $1.8\times10^9$  &\citet{Decarli14}\\ 
SMM~J123707 &2.487 & CO(3$\rightarrow$2) & $1.0\pm0.3$ & 3.20$\pm$0.96 & 11.19$^{a}$ & 4.23$\pm$1.28 & 11.19  & 14& $3.0\times10^9$ &\citet{Bothwell13}\\
SMM~J123606&2.505& CO(3$\rightarrow$2)  & $0.45\pm0.15$ & 1.46$\pm$0.49 & 11.20$^{a}$ & 1.95$\pm$0.64 & 10.30 & 1.81& $1.1\times10^{10}$  &\citet{Bothwell13} \\
GN20&4.055& CO(4$\rightarrow$3)&  1.5$\pm$0.2& 6.01$\pm$0.80 & 11.04$^{b}$ & 15.0$\pm$2.00 & 13.46 & 2622&  $5.7\times10^7$ &\citet{Daddi09} \\
GN20.2a&4.051& CO(4$\rightarrow$3)&0.9$\pm$0.3& 3.61$\pm$1.20 & 11.58$^{b}$ & 9.03$\pm$2.01 & 13.20 & 1441& $6.3\times10^7$  &\citet{Daddi09} \\
HDF850.1  &5.183& CO(2$\rightarrow$1) & $0.148\pm0.057$ & 3.47$\pm$1.34 & --- & 3.08$\pm$1.19  & 12.58 & 346& $8.9\times10^7$  &\citet{Riechers20}\\
GN10&5.303&    CO(1$\rightarrow$0)            &         $0.054\pm0.017$ & 5.24$\pm$1.65 & --- & 4.19$\pm$1.32 & 12.76 &  523&  $8.0\times10^7$
 &\citet{Riechers20} \\
\hline
\hline
\end{tabular}
\caption{Column (1): source ID. Column (2): CO-based spectroscopic redshift. (3): CO(J$\rightarrow$J-1)  transition. (4): velocity integrated CO(J$\rightarrow$J-1) flux. (5): Velocity integrated CO(J$\rightarrow$J-1)  luminosity. (6): Stellar mass by \citet{Hainline2011} or \citet{Tan2014}, denoted with (a) or (b), respectively. The symbol --- denotes that the stellar mass is not available in the literature. (7): Molecular H$_2$ gas estimated assuming a CO-to-H$_2$ conversion factor $\alpha_{\rm CO}=0.8~M_\odot\,({\rm K~km~s}^{-1}~{\rm pc}^2)^{-1}$, typical of starbursts, and an excitation ratio  $r_{\rm J1}=L^\prime_{\rm CO(J\rightarrow J-1)}/L^\prime_{\rm CO(1\rightarrow 0)}=0.9$, $0.6$, and $0.32$ for J$=2$, $3$, and $4$, respectively \citep{Birkin2021}. (8): FIR luminosity. (9): SFR estimated via the relation ${\rm SFR}/(M_\odot/{\rm yr})=9.09\times10^{-11}\,L_{\rm FIR}/L_\odot$ \citep{Kennicutt98}. (10): Depletion timescale $\tau_{\rm dep} = M_{\rm H_2} / {\rm SFR}$. (11): Reference for the CO and FIR observations.}
\label{tab:mol_gas_properties}
\end{table*}

In the last decade, (sub)millimeter surveys have revolutionized our understanding of the formation and evolution of galaxies, by revealing an unexpected population of high-redshift, dust-obscured galaxies that are forming stars at a tremendous rate. Submillimeter  galaxies \citep[SMGs; see the review by][]{Casey14}, first discovered by \citet{Smail97}, have intense star formation rates (SFRs) of a few hundred to several thousand solar masses per year. These dusty starbursts are massive \citep[e.g.,][]{Greve05}, most probably the precursors of present-day ellipticals \citep[e.g.,][]{Ivison13}, and should be excellent tracers of {matter density peaks} and thus protoclusters. In the past few years, several groups have discovered overdensities of dusty star-forming galaxies at redshifts beyond $z\simeq2$ \citep[e.g.,][]{Clements14, Dannerbauer14,Casey15,Clements16,Flores-Cacho16,Hung16,Wang16}. 

Some of these protoclusters, such as the well-known Spiderweb Galaxy protocluster around the powerful radio galaxy MRC1138$+$262 at $z=2.16$, are associated with  galaxy overdensities not only of SMGs but also of low- to medium-mass galaxy populations, such as Lyman $\alpha$ emitters (LAEs) or H$\alpha$ emitters. These studies are complemented by discoveries of a handful of single SMGs physically associated with galaxy overdensities traced at other wavelengths  {\citep[e.g.,][]{Caputi21}}. A notable example is the companion GN20 and GN20.2a SMGs at $z=4.05$, which were discovered by \citet{Daddi09} to lie in a strong galaxy overdensity with a total mass of $\sim10^{14}~M_{\odot}$.

At similar redshifts, \citet{Oteo18} later discovered an extreme protocluster whose core is formed by at least ten dusty star-forming galaxies at $z=4.0$, {located} within a projected area with a size of $\sim140$~kpc. \citet{Miller18} also reported a massive overdensity of 14 gas-rich galaxies at redshift $z=4.3$ within a projected region of $\sim130$~kpc in diameter. At even earlier epochs, there is the protocluster around the SMG AzTEC-3 at $z=5.3$ in the Cosmic Evolution Survey (COSMOS) field \citep{Capak2007}, which contains at least one SMG \citep{Riechers10,Capak11} as well as several LAEs \citep{Guaita2022}.

While generally similar to radio galaxies in terms of their stellar mass, SMGs may have one advantage for
studying the detailed properties of high-redshift clusters: the surface density of bright
SMGs is at least a factor of 10 higher than that of powerful radio galaxies \citep{Reuland03}. Thus, if it can be proven that luminous submillimeter sources do indeed lie in
regions of galaxy overdensities, there is the potential that, in the future, they could be used to substantially increase the number of known high-redshift clusters and protoclusters and thereby facilitate a detailed and statistical study of such large-scale structures.

However, we note that there is a controversy on how reliably SMGs can be used as tracers of large-scale structures \citep{Chapman09,Miller15,Casey16,AlvarezCrespo2021,Gao2022}. By using  large-volume semi-analytic simulations from \citet{Klypin11} and applying abundance matching and analytical prescriptions for galaxies, \citet{Miller15} found  that most of the {strongest} overdensities do not host SMGs. The recent review by \citet{Alberts22} on galaxy protoclusters, focusing on the infrared, discusses the pro and cons and concludes that SMGs could be signposts for overdensities.

{To the best of our knowledge, there are only a few systematic studies that {have investigated} whether SMGs can be used as reliable tracers of galaxy overdensities \citep[e.g.,][]{AlvarezCrespo2021,Gao2022}. In this work} we present a pilot statistical study aimed at exploring the  potential of SMGs to trace galaxy protoclusters. 
To this aim, we use the Poisson probability method \citep[PPM;][]{Castignani2014a} to search for and characterize megaparsec-scale overdensities around a sample of distant $z\sim1.2-5.3$ SMGs, which are possibly associated with large-scale protoclusters. The PPM was primarily introduced and applied to search for high-redshift clusters and groups around $z\sim1-3$ radio galaxies, in particular within the COSMOS survey \citep{Castignani2014b,Castignani2019}, and in this work we apply the method using distant SMGs instead as positional {priors} for our search for protoclusters. The major goal of our work is to shed light on the strongly debated question of whether SMGs are good tracers of distant protoclusters \citep[see, e.g.,][]{Miller15,Casey15}, via testing the efficiency of the PPM as a validation and characterization of protocluster candidates around well-known and spectroscopically confirmed SMGs.

The paper is organized as follows. In Sect.~\ref{sec:sample} we describe the data used in this work: the SMG sample and the photometric redshift catalogs. In Sect.~\ref{sec:PPM} we describe the PPM that we use to search for protoclusters around the SMGs. In Sect.~\ref{sec:results} we present and discuss our results. In Sect.~\ref{sec:conclusions} we draw our conclusions. Throughout this work we adopt a flat $\Lambda \rm$ cold dark matter cosmology with matter density $\Omega_{\rm m} = 0.30$, dark energy density $\Omega_{\Lambda} = 0.70,$ and Hubble constant $h=H_0/100\, \rm km\,s^{-1}\,Mpc^{-1} = 0.70$.

\begin{table*}
\small\addtolength{\tabcolsep}{-3pt}
\centering
 Summary of the SMG properties: coordinates and known overdensities.\\
        \begin{tabular}{l|ccccc}
    \hline
    \hline
Name & RA (J2000) & Dec. (J2000) & $z_{\rm spec}$  & Known overdensity & Reference\\
(1) & (2) & (3) & (4) & (5) & (6) \\
\hline\hline
SMM~J123634& 12:36:34.57 & 62:12:41.0 & 1.225  & &  \\
\hline
ID.03& 12:36:48.63 & 62:12:15.8 & 1.784  && \\
\hline
SMM~J123711& 12:37:11.19 & 62:13:31.2 & 1.995  & yes & \citet{Chapman09} \\
SMM~J123618& 12:36:18.47 & 62:15:51.0 & 1.996  & yes& \citet{Chapman09}\\
SMM~J123712& 12:37:12.12 & 62:13:22.2 & 1.996  &yes& \citet{Chapman09} \\
\hline
ID.19& 12:36:51.60 & 62:12:17.3 & 2.047  && \\
\hline
SMM~J123707& 12:37:07.28 & 62:14:08.6 & 2.487  && \\
\hline
SMM~J123606& 12:36:06.21 & 62:10:24.9 & 2.505  &&\\
\hline
GN20& 12:37:11.90 & 62:22:12.1 & 4.055 & yes&\citet{Daddi09} \\
GN20.2a& 12:37:08.77 & 62:22:01.7 & 4.051  &yes& \citet{Daddi09} \\
\hline
HDF850.1 & 12:36:51.99 & 62:12:25.8 & 5.183  &yes&\citet{Walter12} \\
\hline
GN10& 12:36:33.45 & 62:14:08.9 & 5.303  && \\
\hline
\hline
\end{tabular}
\caption{
Column (1): source ID. Columns (2,3): J2000.0 right ascension and declination. Column (4): CO-based spectroscopic redshift. Column (5): known overdensity flag. Column (6): Reference for the overdensity around the SMG.\\
}
\label{tab:known_overdensities}
\end{table*}

\begin{figure}[htbp]
\begin{center}
\includegraphics[width=9cm]{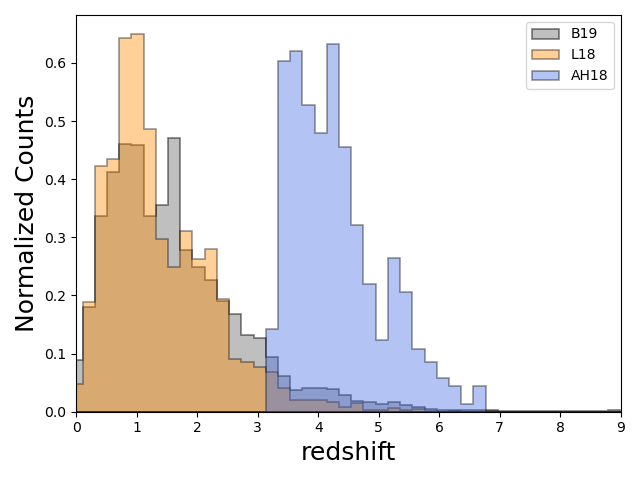}
\caption{Photometric redshift distributions of the B19 \citep{Barro19}, L18 \citep{Liu18}, and AH18 \citep{Arrabal18}  catalogs in the GOODS-N field.}
\label{fig:red_dist}
\end{center}
\end{figure}

\begin{figure*}[htbp]
\begin{center}
\includegraphics[trim={2cm 1cm 1cm 0},width=16cm]{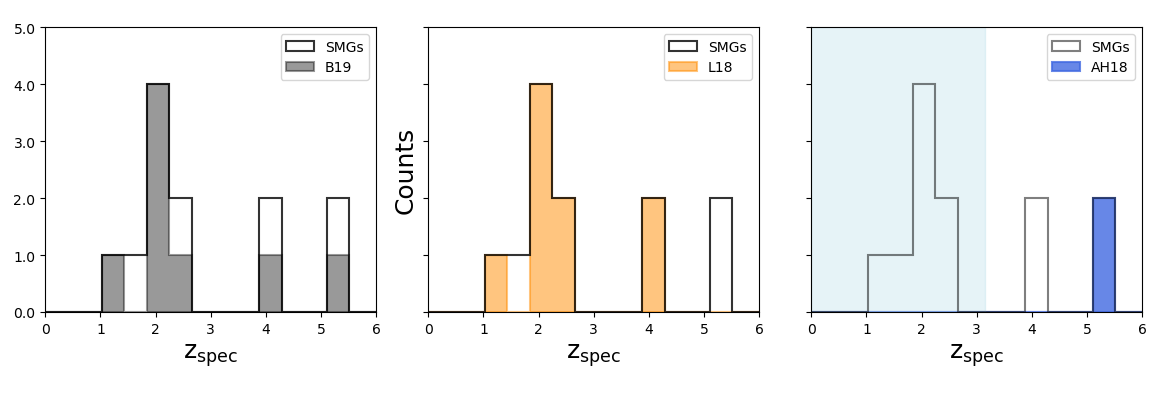}
\caption{Redshift distribution of the SMGs. The associated overdensities found using the B19 (left), L18 (center), and AH18 (right) photo-$z$ catalogs are highlighted with different colors. The shaded region in the right histogram indicates that there are no sources 
at $z<3.35$ in the AH18 catalog.}
\label{fig:redshift_histo_PPMoverdensities}
\end{center}
\end{figure*}

\begin{figure*}
\centering
\vspace{1cm}
\includegraphics[trim={0cm 0cm 0cm 0cm},width=0.49\textwidth]{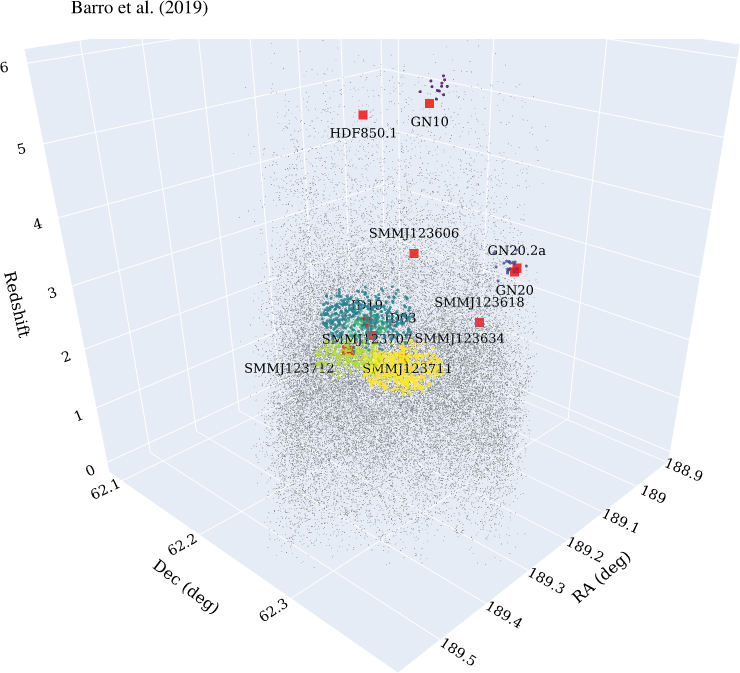}
\vspace{1cm}
\includegraphics[trim={0cm 0cm 0cm  0cm},width=0.49\textwidth]{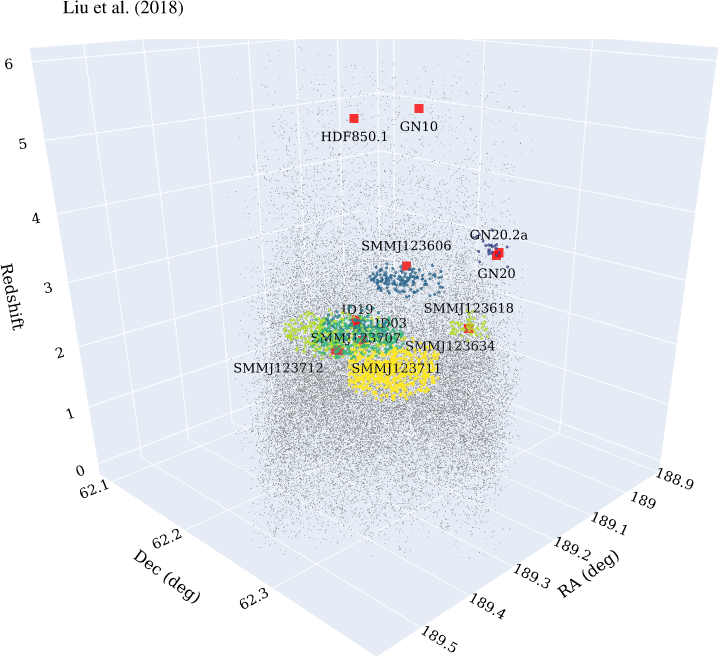}\\
\includegraphics[trim={0cm 0cm 0cm 0cm},width=0.49\textwidth]{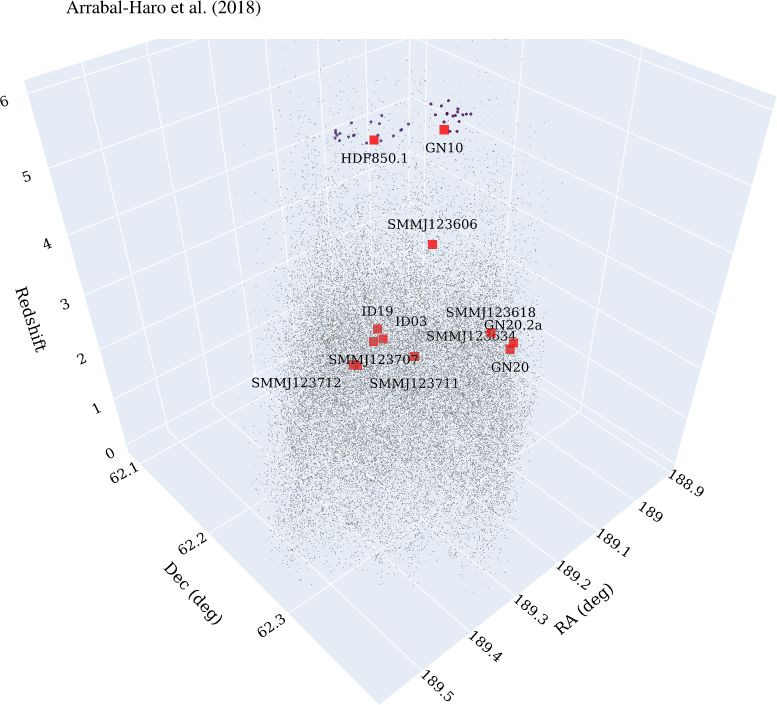}
\caption{Distribution of the SMGs and associated protoclusters in the RA, Dec., and redshift space when the B19, L18, and AH18 photometric redshift catalogs are used. Small gray dots are sources from the L18 catalog, while red squares show the location of the SMGs in our sample. Colored clouds show the members of the overdensities as identified with the PPM. Protocluster members are color coded in yellow, green, and blue according to their increasing redshift. Interactive 3D plots for all
three photometric redshift catalogs are available electronically.}
\label{fig:3Dplots}
\end{figure*}

\begin{figure}[htbp]
\centering
\includegraphics[trim={2cm 1cm 3cm 0cm},width=0.4\textwidth]{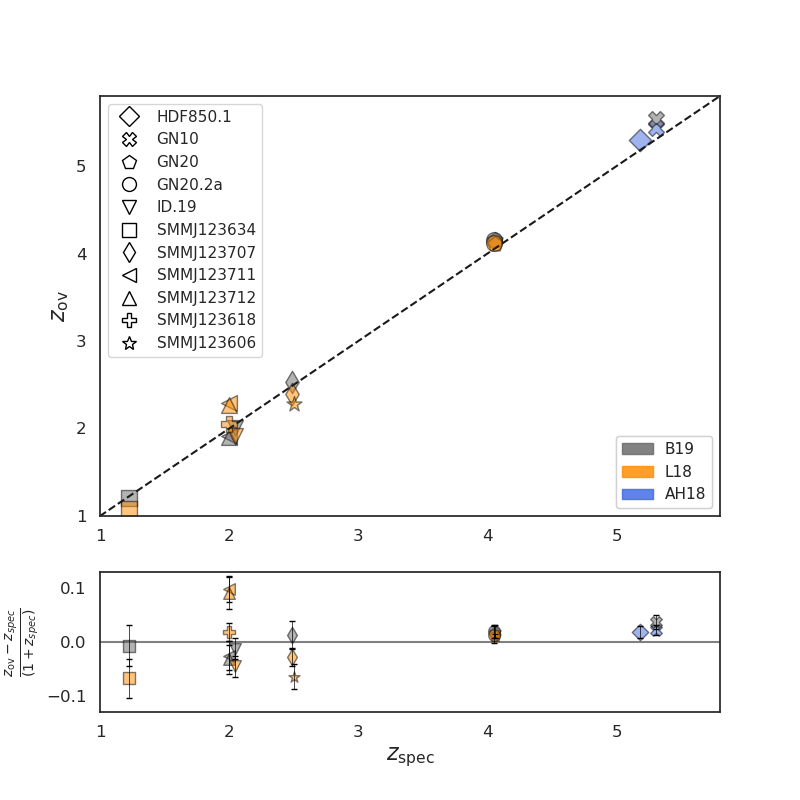}
\caption{Top panel: Overdensity redshift measured by the PPM ($z_{\rm ov}$) as a function of the spectroscopic redshift of the SMG ($z_{\rm spec}$). We refer to the legends at the top left and bottom right for the symbols and the color code used for the different SMGs and photo-$z$ catalogs. Bottom panel: Redshift residuals as a function of $z_{\rm spec}$.}
\label{fig:redshifts_accuracy}
\end{figure}

\section{Data}\label{sec:sample}
In this section we present the main data from the literature that we use in our study. We first describe our sample of SMGs, and then the  different photometric redshift catalogs that we use to search for galaxy overdensities around the SMGs themselves. We emphasize that with the PPM we will be able to search for galaxy overdensities at megaparsec scales, which is the typical size of protocluster cores, whereas the overall extent of galaxy protoclusters could be up to 10-20~Mpc  \citep[see, e.g.,][]{Muldrew15,Casey15}.

\subsection{The submillimeter galaxies}\label{sec:SMGprop}
\subsubsection{The selection of the SMG sample}
We searched in the literature for SMGs in the GOODS-N (Great Observatories Origins Deep Survey) field\footnote{\url{https://www.stsci.edu/science/goods/}}, with spectroscopic redshifts obtained via CO lines, which is {the most} 
reliable and accurate methods for determining the redshift of these dusty starbursts. This search resulted in 17 SMGs.  However, as we want to search for galaxy overdensities with the PPM, which makes uses of photometric redshifts of galaxies in the field of the SMGs, we removed five sources from the sample of 17 as they fall outside the range of RA and Dec. covered by the photo-$z$ catalogs that are currently available for the GOODS-N field.  Our final SMG sample thus consists of 12 gas-rich, star-forming, and bright far-infrared (FIR) sources, as further discussed in the following. In Table~\ref{tab:mol_gas_properties} we report the main properties of the 12 SMGs of this work, including their redshift, stellar and molecular gas masses,  CO and FIR luminosities, SFR, and depletion time $\tau_{\rm dep} = M_{\rm H_2}/{\rm SFR}$.
We note that two SMGs, namely, ID.03 and ID.19, have not been selected through submillimeter surveys but they were found directly through a CO blind survey \citep{Decarli14}. 

\subsubsection{Molecular gas and star formation}\label{sec:molecular_gas_prop}

Molecular gas masses of the SMGs are high, ranging between $(1.3-15)\times 10^{10}~ M_{\odot}$. They come from flux measurements from the literature, and were estimated as follows. For all 12 SMGs we first searched in the literature for CO(J$\rightarrow$J-1) line flux measurements, where J is a positive integer denoting the total angular momentum. When multiple lines were available for a given SMG, we gave preference to the lowest-J transition available. {This is because lower-J CO transitions require} lower gas densities  and temperatures \citep[e.g.,][]{Liu2021} and are thus preferred when estimating the total cold molecular gas masses. We then converted CO(J$\rightarrow$J-1) fluxes into velocity integrated luminosities $L^\prime_{\rm CO(J\rightarrow J-1)}$, via Eq.~3 of \citet{Solomon_VandenBout2005},  and then into {molecular} gas masses. 
In order to have homogeneous gas mass estimates we used a CO-to-H$_2$ conversion factor $\rm \alpha_{CO} = 0.8~ M_{\odot}~ (K~km~s^{-1}~pc^{2})^{-1}$, typical of starbursts, and excitation ratios
$r_{\rm J1}=L^\prime_{\rm CO(J\rightarrow J-1)}/L^\prime_{\rm CO(1\rightarrow 0)}=0.9$, $0.6$, and $0.32$ for J$=2$, $3$, and $4$, respectively \citep{Birkin2021}.

It is worth noticing that the SMGs in our sample not only have high molecular gas masses but they also have high FIR luminosities, with a median value of $\rm log(L_{\rm FIR}/L_{\odot})\simeq12.6$, typical of ultra-luminous infrared galaxies (ULIRGs; see Table~\ref{tab:mol_gas_properties}). The sources in our sample are thus bright SMGs, similar to those originally discovered with bolometer cameras mounted on single-dish telescopes \citep[e.g.,][]{Smail97,Hughes1998,Barger1998}.

The FIR luminosities were then converted into SFRs using the relation ${\rm SFR}/(M_\odot/{\rm yr})=9.09\times10^{-11}
L_{\rm FIR}/L_\odot$ by \citet{Kennicutt98}. Hence, our SMGs are characterized by strong ongoing star formation activity, with a median of {${\rm SFR} \simeq 340~ M_\odot/{\rm yr}$.} In particular, {the} GN20--GN20.2a pair of SMGs have the highest SFRs, exceeding $1000~M_\odot$/yr, and FIR luminosities largely exceeding the value of $\rm log(L_{FIR}/L_{\odot})=13$, which is typical of hyper-luminous infrared galaxies (HyLIRGs; see Sect.~\ref{sec:SMG_env_general}). These high levels of star formation activity in the SMGs are ultimately sustained by the large {molecular} gas reservoirs, which are consumed quite efficiently, which is typical of starbursts. Our SMGs have indeed short depletion timescales $\tau_{\rm dep}=M_{\rm H_{2}}/{\rm SFR}\simeq0.1$~Gyr (median).

\subsubsection{Stellar masses}
We found stellar masses for only seven SMGs in the literature \citep{Hainline2011,Tan2014}, in the range $\log(M_\star/M_\odot)\simeq10.5-11.2$ (see Table~\ref{tab:mol_gas_properties}). Many SMGs are indeed undetected in the optical wavelengths, as for example it is the case for GN10 and HDF850.1, which are completely dust-obscured \citep{Daddi09, Walter12,Calvi21}, so that it is difficult to estimate their stellar masses. 
Altogether, the SMGs in our sample have high stellar masses and star formation, and are thus likely the progenitors of local ellipticals
\citep{Michalowski2010,Fu2013}.

\subsubsection{The SMGs and their environments: General considerations}\label{sec:SMG_env_general}

Previous studies found evidence of protoclusters for six SMGs out of the 12 in our sample, as discussed in the following. SMM~J123711, SMM~J12618, and SMM~J123712 were confirmed by \citet{Bothwell13} as bright submillimeter emitters, with similar properties in terms of gas masses $M_{\rm H_2}\simeq(3-5-5.5)\times10^{10}~M_\odot$ and FIR luminosities, $L_{\rm FIR}/L_\odot=12.4-12.9$, typical of ULIRGs. They also have the same redshifts $z=1.995-1.996$ and relatively small pairwise projected separations of $\sim3.3$~Mpc. At the location of these SMGs, \citet{Chapman09} reported the existence of a $z=1.99$ protocluster, discovered as a spectroscopic overdensity of $\sim100$ SMGs across 800~arcmin$^2$. Therefore, we believe all three SMGs in our sample belong to the same overdense large-scale structure.

GN20 and GN20.2a were discovered as bright distant SMGs by \citet{Daddi09}. They have similarly high gas masses $M_{\rm H_2}=(0.9-1.5)\times10^{11}~M_\odot$ and FIR luminosities, $L_{\rm FIR}/L_\odot=13.20-13.46$, typical of HyLIRGs  {\citep[e.g.,][]{RowanRobinson2000,Neri2020}}, which are often associated with infall/outflows, mergers, and mutual interactions. They are thus possibly interacting as indeed they are separated by just 24~arcsec (i.e., 167~kpc) in projection, while being at the same redshift $z=4.05$. Furthermore, \citet{Daddi09} found a strong overdensity of B-band dropouts and IRAC (Infrared Array Camera on the
Spitzer Space Telescope) -selected massive $z>3.5$ galaxies, appearing to be centered on the two SMGs GN20 and GN20.2a, which suggests the presence of a $z=4.05$ protocluster. 

Similarly, the $z=5.2$ SMG HDF850.1 in our sample is embedded in an overdensity of galaxies with spectroscopic redshifts that agree well with that of HDF850.1 \citep{Walter12}. \citet{Calvi21} showed none of their 23 spectroscopically confirmed members are SMGs. They are instead in majority LAEs or Lyman break galaxies (LBGs).

To the best of our knowledge, these are the only known (candidate) protoclusters around SMGs in the GOOD-N field. The remaining six SMGs in our sample have no
unambiguous environmental analysis in the literature and have pairwise large separation both in redshift and in projection. In this work we investigate the large-scale environments of all 12 SMGs using the PPM, considering each of them separately. However, given the considerations outlined above, we then discuss the detected overdensities associated with the SMGs GN20 and GN20.2a altogether, and similarly for the SMGs SMM~J123711, SMM~J12618, and SMM~J123712.

In Table~\ref{tab:known_overdensities} we provide a summary of the basic properties of the 12 SMGs, which are grouped in those cases where multiple SMGs are close to each other and likely belong to same overdensity, as discussed above.

\subsection{Photometric redshift catalogs}\label{sec:photoz_catalogs}
In this section we give a brief overview of the three photometric redshift catalogs we used to search for galaxy protoclusters around the SMGs.
These catalogs were built and released by \citet[][]{Arrabal18}, \citet[][]{Barro19}, and \citet[][]{Liu18}, hereafter denoted as AH18, B19, and L18, respectively, to which we refer for a complete complete discussion of the sample built-up and detailed properties. As outlined below, the three catalogs contain sources that were selected with different criteria and techniques, as well as deep multiwavelength photometry from the infrared to the ultra-violet. They thus contain complementary information about the high-$z$ galaxy population in the GOOD-N field and are therefore optimal to search for and characterize distant protoclusters limiting the possible biases that could arise by using instead a single multiband photometric catalog.

\subsubsection{Barro et al. (2019)}This is the largest photo-$z$ catalog we used. The sources were selected by the authors from the CANDELS (Cosmic Assembly Near-infrared Deep Extragalactic Legacy Survey) survey \citep{Grogin11,Koekemoer11} covering the GOODS-N field.      
      The selection was done using the WFC3/F160W filters ($H$-band) down to a magnitude limit of $H\sim$28~mag. The selection also combined data from the deep and wide-field observations of the CANDELS program, from the ultraviolet to the FIR, over a total area of 171~arcmin$^{2}$. The released catalog contains 35445 sources, of which 22670 have photometric redshifts in the range $1.0<z<5.5$. The highest and nearly uniform accuracy $\sigma(\frac{\delta z}{1+z_{\rm spec}})\lesssim0.01$  of the photometric redshifts is found in the redshift range $z\sim0-2.5$. Altogether, the median redshift of the sources in the catalog is $z=1.43^{+1.22}_{-0.85}$, where the uncertainties correspond to the 68\% confidence interval around the median. The B19 catalog thus allows us to potentially search for overdensities around all SMGs in our sample.

\subsubsection{Liu et al. (2018)}This catalog includes photometry for 3306 super-deblended {\it Herschel} sources in the GOODS-N field, up to $z\sim6$. The sample was built by the authors using a technique that allowed them to process low-resolution FIR images, thus limiting source confusion. Specifically, they used positional priors based on source positions from higher-resolution infrared and radio observations with {\it Spitzer}/MIPS (Multiband Imaging Photometer for Spitzer) at 24$\mu$m and VLA (Karl Guthe Jansky Very Large Array) at 1.4~GHz. These positional priors allowed the authors to properly fit the FIR/submillimeter data, and thus build an unblended source catalog with reliable photometry in both FIR and (sub)millimeter bands. \\

\subsubsection{Arrabal-Haro et al. (2018)} This photometric catalog consists of 1558 high-$z$ galaxies selected in the ultraviolet (528 LAEs and 1030 LBGs), which were  selected with criteria based on color excesses and spectral energy distribution fits. In more detail, the authors carried out a systematic search for LAEs and LBGs between $z\sim3.35-6.8$ by using 25 medium-band filters (full width at half maximum~$\sim17$~nm; from 500 to 941~nm) of the Survey for High-$z$ Absorption Red and Dead Sources \citep[SHARDS;][]{Perez13}. This survey was conducted with the instrument OSIRIS (Optical System for Imaging and low-Intermediate-Resolution Integrated Spectroscopy) \citep{Cepa2010} at the GranTeCan telescope, covering an area of $\sim 130$ arcmin$^{2}$ of the GOODS-N field.     
      Figure~\ref{fig:red_dist} displays the redshift distribution of galaxies in the three photometric catalogs. It shows the complementarity in terms of redshift range of the AH18 catalog with respect to the other two. While both B19 and L18 samples have similar redshift distributions peaking around $z\sim1$, AH18 sources populate the highest redshifts. \\

\begin{table*}[h!]
\small\addtolength{\tabcolsep}{-3pt}
\centering
\begin{tabular}{c|cccccccccc}
\hline\hline
 Name &  $z_{\rm spec}$ & $({\rm R.A.})_{\rm ov}$ & $({\rm Dec.})_{\rm ov}$ & $\theta_{\rm ov}$ & $z_{\rm ov}$ & significance &  $N_{\rm selected}$  & $\mathcal{R}_{\rm PPM}$ & $\mathcal{R}_{\mathit{w}}$  &  catalog \\
 (1) & (2) & (3) & (4) & (5) & (6) & (7) & (8) & (9) & (10) & (11) \\ 
 \hline
\hline
\multirow{4}{*}{SMM~J123634} & \multirow{4}{*}{1.2245} & 12:36:48.7 & 62:11:32.5 & 120.4$''$ & 1.12$\pm$0.09   & 2.0$\sigma$  &  197~(164.6) & {120$''$}  &  {136$''$} & B19 \\
 &  & 12:36:45.7 & 62:11:31.2 & 104.4$''$ & 1.02$\pm$0.07   & 2.1$\sigma$  &  214~(185.4) & {112$''$}  &  {156$''$} & B19 \\
 &  & 12:36:33.5  & 62:14:10.4   & 89.7$''$ & 1.32$\pm$0.08   & 2.0$\sigma$  &  22~(11.4) & {134$''$}  &  {67$''$} & L18 \\
&  &   12:36:20.3 & 62:11:24.3 & 125.7$''$ & 1.08$\pm$0.09   & 2.1$\sigma$  &  29~(20.8) & {134$''$}  &  {69$''$} & L18 \\
  \hline
  \multirow{2}{*}{\small ID.03} &  \multirow{2}{*}{\small 1.7844} & 
12:37:02.5 & 62:12:00.8  & 98.2$''$ & 1.28$\pm$0.08   & 3.7$\sigma$  &  292~(232.9) & {104$''$}  &  {184$''$} & B19~($\ast$) \\
 &    & 12:37:00.3 & 62:11:53.6
   & 84.4$''$ & 1.42$\pm$0.08   & 2.2$\sigma$  &  8~(3.2) & {104$''$}  &  {53$''$} & L18~($\ast$) \\
   \hline
   \multirow{2}{*}{SMM~J123711} & \multirow{2}{*}{1.9951} & 12:37:04.8 & 62:12: 24.2 & 80.5$''$ & 1.92$\pm$0.08   & 2.2$\sigma$  &  153~(126.5) & {85$''$}  &  {233$''$} & B19 \\
 &  & 12:37:02.5 & 62:14:54.9  & 103.5$''$ & 2.29$\pm$0.07   & 2.7$\sigma$  &  8~(2.6) & {104$''$}  &  {68$''$} & L18 \\
 \hline
 SMM~J123618 & 1.9964 & 12:36:21.7 & 62:16:20.9  &  37.4$''$ & 2.04$\pm$0.05   & 3.0$\sigma$  &  10~(3.1) & {60$''$}  &  {100$''$} & L18 \\
  \hline
 \multirow{2}{*}{SMM~J123712} &\multirow{2}{*}{1.9964} &  12:37:05.7 & 62:12:15.3  & 80.4$''$ & 1.90$\pm$0.08   & 2.1$\sigma$  &  152~(128.5) & {85$''$}  &  {116$''$} & B19 \\
 &  & 12:37:02.3 & 62:15:01.0  & 120.2$''$ & 2.27$\pm$0.09   & 2.3$\sigma$  &  15~(7.5) & {134$''$}  &  {102$''$} & L18 \\
 \hline
 \multirow{2}{*}{ID.19} & \multirow{2}{*}{2.047} & 12:36:59.1  & 62:11:54.9  & 56.9$''$ & 2.00$\pm$0.06 & 2.0$\sigma$  &  54~(41.1) & 60$''$  &  90$''$ & B19~(+) \\
 &  & 12:36:37.8 &  62:11:10.4  &  117.5$''$ & 1.90$\pm$0.06   & 2.0$\sigma$  &  7~(2.9) & 134$''$  &  66$''$ & L18~(+) \\
 \hline
 \multirow{3}{*}{SMM~J123707} & \multirow{3}{*}{2.4870} & 12:37:13.9  & 62:12:44.1  & 96.3$''$ & 2.40$\pm$0.09   & 2.5$\sigma$  &  168~(134.9) & {120$''$}  &  {103$''$} & B19 \\
 &  & 12:37:10.6  &  62:12:50.2  & 81.9$''$ & 2.65$\pm$0.09   & 3.1$\sigma$  &  197~(158.1) & {120$''$}  &  {140$''$} & B19 \\
 &  &  12:37:02.9  & 62:15:02.3   &  61.8$''$ & 2.38$\pm$0.07  & 2.5$\sigma$  &  8~(2.9) & {85$''$}  &  {86$''$} & L18 \\
 \hline
 SMM~J123606 & 2.5054 & 12:36:09.5 & 62:08:53.7  & 94.1$''$ & 2.27$\pm$0.08   & 2.6$\sigma$  &  16~(8.1) & {120$''$}  &  {102$''$} & L18 \\
 \hline
 \multirow{2}{*}{GN20}  &\multirow{2}{*}{4.055} & 12:37:11.9  & 62:22:03.0 & 9.1$''$ & 4.15$\pm$0.06   & 2.9$\sigma$  &  20~(9.7) & {60$''$}  &  {102$''$} & B19 \\
        &          & 12:37:10.6 & 62:22:03.0 & 12.9$''$ & 4.11$\pm$0.02   & 5.4$\sigma$  &  5~(0.06) & {60$''$}  &  {102$''$} & L18 \\
\hline
 \multirow{2}{*}{GN20.2a}       &          \multirow{2}{*}{4.051} & 12:37:12.7  &  62:22:01.7  & 27.4$''$ & $4.15\pm0.06$  & 2.0$\sigma$  &  14~(13.6) & 60$''$  & 122$''$ & B19~(+) \\
        &                & 12:37:10.1 & 62:22:01.7  & 9.1$''$ & 4.12$\pm$0.08   & 5.0$\sigma$  &  6~(0.12) & {85$''$}  &  {81$''$} & L18 \\
\hline
\multirow{2}{*}{\small HDF850.1} & \multirow{2}{*}{\small 5.183} & 12:37:03.7 & 62:11:55.1  & 87.5$''$ & 5.29$\pm$0.07   & 2.8$\sigma$  &  24~(13.1) & {104$''$}  &  {137$''$} & AH18 \\
   &     & 12:36:56.6  & 62:13:52.3   & 92.4$''$ & 5.87$\pm$0.07   & 3.0$\sigma$  &  15~(5.2) & {120$''$}  &  {121$''$} & B19~($\ast$) \\
\hline

\multirow{2}{*}{GN10} & \multirow{2}{*}{5.303}  & 12:36:29.0 & 62:13:58.5  & 32.8$''$ & 5.44$\pm$0.06   & 2.1$\sigma$  &  14~(7.7) & {85$''$}  &  {162$''$} & AH18 \\             
 &   &  12:36:30.4 & 62:14:19.3 & 23.4$''$ & 5.54$\pm$0.08   & 2.7$\sigma$  &  12~(4.4) & {60$''$}  &  {164$''$} & B19 \\        
 \hline\hline
 \end{tabular}

\caption{Properties of the megaparsec-scale overdensities around the SMGs:  (1-2) galaxy ID and spectroscopic redshift; (3-4) J2000 projected space coordinates of the overdensity (ov) peak as found by the wavelet transform; (5) projected separation between the coordinates in columns (3-4) and those of the SMG; (6) overdensity  redshift and (7) significance as found by the PPM;  (8) number of sources selected by the PPM to detect the overdensity, while between parentheses there is the corresponding average number of sources in the survey within the overdensity area and the redshift bin of $\overline{\Delta z}\simeq0.3$ centered around $z_{\rm ov}$; (9-10)  maximum radius within which the overdensity is detected as found by the PPM and by the wavelet transform; (11) adopted photometric redshift catalog.\\
{\bf Notes.} Sources denoted with an asterisk ($\ast$) in column (11) are those where the estimated redshift of the overdensity is marginally consistent, that is, $|z_{\rm spec}-z_{\rm ov}|\gtrsim0.3$, with that of the SMG. Sources denoted with the symbol (+) in column (11) are instead those where an overdensity is detected when choosing a lower parameter $\overline{\Delta z}=0.2$ than that used for the others (see Sect.~\ref{sec:PPM}).}

\label{tab:cluster_properties}
\end{table*}

\section{The Poisson probability method}\label{sec:PPM}
The PPM searches for high-$z$ megaparsec-scale overdensities of galaxies around a given target. It is based on a theory defined on the ensemble of the photometric redshift realizations of the galaxies in the field.  Through the use of a solid positional prior and an accurate photometric redshift sampling, the  PPM partially overcomes the limitations deriving from low number-count statistics and shot-noise fluctuations, which  are particularly relevant in the high-$z$ universe such as in the case of protoclusters. 
More specifically, the PPM method uses photometric redshifts of galaxies to search for overdensities around each target (each SMG, in this work)  along the line of sight. To this aim the PPM adopts an accurate sampling of the photometric redshift information to the detriment of a less sophisticated tessellation of the projected space, which is performed in terms of concentric annuli centered around each target.
We refer to our previous studies for a detailed description of the method \citep{Castignani2014a}, its wavelet based extension \citep[$w$PPM;][]{Castignani2019}, and the applications \citep{Castignani2014b,Castignani2019}.
We summarize below the basic steps of the method. 

First, we tessellated the projected space with a circle centered at the coordinates of the SMG and a number of consecutive adjacent annuli. The annuli and the central circle have an equal area {of 3.14~arcmin$^2$}. In particular, the circle has a radius of 60~arcsec, which corresponds to physical scales in the range $\simeq$(0.4-0.5)~Mpc for the SMGs in our sample.

Second, for each region {of the tessellation (the central circle and the consecutive annuli)}, we counted the number of sources with photometric redshifts within an interval of length $\Delta z$ and  centered at the centroid redshift, $z_{\rm centroid}$. We refer to Sect. \ref{sec:results} for a description of the photometric redshift catalog considered. The parameters $\Delta z$ and $z_{\rm centroid}$ uniformly span a grid of values that reflect the photometric redshift uncertainties and correspond to the redshift range of our interest, respectively.

Third, for each pair ($z_{\rm centroid};\Delta z$), we mapped the galaxy number counts into an overdensity measure that relies on the significance, based on Poisson statistics, that the null hypothesis of no clustering is rejected at the location of the  SMG. {For this,}  we estimated the mean field density using a rectangular control region concentric with the GOODS-N field and with a subtended area of 50.3~arcmin$^2$. The corresponding PPM plots {(Appendix~\ref{app:ppmplots})} show the overdensity patterns at different redshifts, where we further applied a Gaussian filter to eliminate high-frequency noisy patterns. We chose a conservative smoothing scale of 0.02 in redshift, on the order of or smaller than typical photometric redshift uncertainties. Overdensities along the line of sight of the SMG were searched for by fixing the redshift bin $\overline{\Delta z}=0.3,$ which roughly maximizes the overdensity significance as it corresponds to a $\sim(3-5)\sigma$ interval for high-$z$ sources with typical photometric redshift uncertainties of $0.02(1+z)$. 

Fourth, at fixed $\overline{\Delta z}$, a peak-finding algorithm was applied to the PPM plot to search for overdensities around the SMGs. This procedure relies on the Morse theory. It allows us to estimate the significance of the overdensity, its redshift $z_{\rm ov}$, and the projected radius $\mathcal{R}_{\rm PPM}$, which is the maximum separation from the SMG up to which the overdensity is detected, while $N_{\rm selected}$ is the number of sources that contribute to the overdensity, within the radius $\mathcal{R}_{\rm PPM}$.
Any overdensity that is located at a redshift consistent with that of the SMG is associated with the galaxy, as done in \citet[][]{Castignani2014b,Castignani2019}. Multiple overdensity associations are not excluded. 

Fifth, similarly to \citet{Castignani2019}, we then applied a wavelet transform to further characterize the overdensity in the projected space. In particular, we provide the projected miscentering, $\theta_{\rm ov}$, of the overdensity peak, as found with the wavelet transform, with respect to the SMG position in the sky. We also derived a second estimate for the overdensity size, $\mathcal{R}_w$, which is the projected radius of the overdensity as found with the wavelet transform. 

Hereafter we refer to the PPM cluster finder together with its wavelet extension as $w$PPM.\\

\section{Results}
\label{sec:results}

We ran the $w$PPM as described in Sect.~\ref{sec:PPM} on the three photometric catalogs presented in Sect.~\ref{sec:photoz_catalogs}. As illustrated in the redshift distributions of Fig.~\ref{fig:red_dist}, both L18 and B19 catalogs span the full redshift range of our SMGs and we are thus potentially able to detect protoclusters around them using these photo-$z$ catalogs. On the other hand, only four SMGs at the highest redshifts $z\sim4.05-5.30$ out of 12 fall within the redshift range of the AH18 catalog, namely,  GN20, GN20.2a, HDF850.1, and GN10.  

Table~\ref{tab:cluster_properties} summarizes the properties of the overdensities (ov) detected with the $w$PPM along the line of sight of each SMG. In particular, we report the projected coordinates of the overdensity peak, the projected miscentering $\theta_{\rm ov}$ with respect to the SMG coordinates, the estimated redshift of the overdensity ($z_{\rm ov}$), the overdensity detection significance, as well as estimates for the overdensity richness ($N_{\rm selected}$) and size. In Appendix~\ref{app:ppmplots} we report the density maps and PPM plots for all the overdensities reported in Table~\ref{tab:cluster_properties}.

\subsection{Global properties of the detected overdensities}

Figure~\ref{fig:redshift_histo_PPMoverdensities} displays the redshift distribution of the 12 SMGs in our sample, where we highlight with different colors the overdensities detected using the B19, L18, and AH18 catalogs, separately. As the redshift coverage of three photo-$z$ catalogs is not the same, as well as the corresponding sample selections, we do not expect to find the same overdensities in all three catalogs. Comparing the overdensity redshifts for the three different photo-$z$ catalogs we note that for both L18 and B19 catalogs the PPM is  effective in detecting overdensities over the full redshift range spanned by the SMGs, while with the AH18 we detect overdensities only around the highest-redshift SMGs.

As shown in Table~\ref{tab:cluster_properties}, there are 11 SMGs that are associated with overdensities in at least one photo-$z$ catalog.
In particular, in five cases, overdensities are found in B19 and L18 catalogs, which cover the full redshift range of our SMGs. These L18 and B19 overdensities are all at $z<4$,  with the exception of GN20. Similarly, all our $z>4$ SMGs appear to be in overdensities in at least two photo-$z$ catalogs.  

Overdensities are also found around the remaining SMG ID.03 with both the B19 and L18 catalogs, but at lower redshifts than that of the SMG, and so the association with the SMG is uncertain. Therefore, unless specified otherwise, in the following we do not consider these overdensities around ID.03.

In three cases we found overdensities around the SMGs only when choosing a slightly lower value,  $\overline{\Delta z}=0.2$, than that used for the other overdensities (i.e., $\overline{\Delta z}=0.3$; see Sect.~\ref{sec:PPM}). These three overdensities are those around ID.19 (in the case of both B19 and L18 catalogs) and GN20.2a (B19), and for our analysis we consider them altogether with the others. However, for the sake of clarity we highlight them in Table~\ref{tab:cluster_properties}. 

In Fig.~\ref{fig:3Dplots} we show the 3D distribution of the 12 SMGs in our sample as well as that of the galaxies of the three photometric redshift catalogs (B19, AH18, and L18), separately, which are used in this work to search for protoclusters. Fiducial overdensity members as found by the $w$PPM  are highlighted (see Sects.~\ref{sec:PPM} and \ref{sec:structural_properties_PCs} for further details). We grouped overdensities, and their members, around mutually close SMGs, as in Table~\ref{tab:known_overdensities} and discussed in Sect.~\ref{sec:SMG_env_general}.
  The interactive version of Fig.~\ref{fig:3Dplots} is available electronically. 
  In the following we further discuss our detections when considering separately the three redshift catalogs.

\subsubsection{Barro et al. (2019)}
For B19 we have {eight} detected overdensities around the SMGs, including GN20, GN10, and possibly HDF850.1 at the highest redshifts $z\sim4-5$, even though for the last one the association of the SMG with the overdensity is uncertain and not reported in Fig.~\ref{fig:redshift_histo_PPMoverdensities}. As highlighted in Table~\ref{tab:cluster_properties}, this is because the  overdensity is detected at a higher redshift, $z_{\rm ov} = 5.87\pm0.07,$ than the spectroscopic one of the SMG, $z_{\rm spec} = 5.183$. A similar uncertain association is found for ID.03 at $z = 1.7844$, discussed above, while the overdensity redshift is $z_{\rm ov} = 1.29\pm0.08$ and $z_{\rm ov} = 1.43\pm0.08$ when the B19 and L18 catalogs are used, respectively.

\subsubsection{Liu et al. (2018)}
When using the L18 catalog {nine} SMGs are found in overdensities up to $z\sim4$ (GN20--GN20.2a). However, we miss two overdensities in L18 around the most distant SMGs GN10 and HDF850.1, which are however detected with the other two photometric redshift catalogs (B19, AH18). At lower redshifts, we note that the catalog has completeness issues below $z=2$ (see Fig.~4 in L18). Despite this, all SMGs at $z<2$ appear to be in L18 overdensities, with some caveats, which we discuss below. The source SMM~J123711 at $z=1.9951$ is associated with a $\simeq3\sigma$  overdensity at $z_{\rm ov}=2.29\pm0.07$; therefore, the two redshifts are marginally consistent with each other, $|z_{\rm ov}-z_{\rm spec}|/(1+z_{\rm spec}) = 0.10$.
Similarly, in the case of ID.03, discussed just above, the association in redshift between the SMG and the overdensity is uncertain, with $|z_{\rm ov}-z_{\rm spec}|/(1+z_{\rm spec}) = 0.14$.

\subsubsection{Arrabal-Haro et al. (2018)}
Out of the four distant SMGs that fall in the $z\sim3.35-6.8$ redshift range of AH18, overdensities are detected around HDF850.1 and GN10, separately.
No overdensity is detected for the other high-$z$ SMGs GN20 and GN20.2a using AH18. Nonetheless, \citet{Arrabal18} reported a redshift peak of their ultraviolet-selected galaxies at the redshift of GN20 and GN20.2a. Furthermore, as discussed in Sect.~\ref{sec:SMG_env_general}, \citet{Daddi09} found a spectroscopic overdensity around the two companion SMGs at $z=4.05$ and we similarly find high signal-to-noise overdensities when using the L18 and B19 catalogs (see Table~\ref{tab:cluster_properties}). 
We double-checked the AH18 catalog and found that the SMGs GN20 and GN20.2a are located at the edge of the footprint of the photo-$z$ catalog, which prevents us from detecting the overdensity, given the circular PPM tessellation of the projected space. A more accurate treatment of edge effects will be implemented in a forthcoming upgrade of the PPM.

\subsection{Multiple associations}\label{sec:multiple_associations}
As listed in Table~\ref{tab:cluster_properties}, for a given photo-$z$ catalog, most SMGs have at most one associated overdensity. However, in two cases, namely the SMGs SMM~J123634 and SMM~J123707, there are multiple overdensities that are detected when using the L18 and B19 catalogs. These correspond to distinct overdensity peaks in the PPM plots at redshifts that are formally consistent with that of the SMG (see Appendix~\ref{app:ppmplots}, where the PPM plots are reported). Similar multiple associations were discussed in our previous studies, \citet{Castignani2014b,Castignani2019}, where we looked for overdensities around distant radio galaxies. Altogether, for a given photometric redshift catalog (B19 or L18), the multiple overdensities reported for SMM~J123634 and SMM~J123707, separately, have similar properties in terms of significance,  size, projected coordinates, and thus miscentering $\theta_{\rm ov}$ (see the density maps in Appendix~\ref{app:ppmplots}). These aspects suggest that these multiple peaks correspond to the same overdensity that is fragmented at different redshifts by the PPM procedure. This behavior is not uncommon for distant (proto)cluster finders, especially in the regime of low number counts.

\subsubsection{Overdensity redshift accuracy}
We now compare the SMG spectroscopic redshifts with those of the overdensities detected by the $w$PPM using photometric redshifts of galaxies.
Figure~\ref{fig:redshifts_accuracy} (top) shows the redshifts of the overdensities, estimated with the PPM using the B18, L18, and AH18 photo-$z$ catalogs, which are plotted against the spectroscopic redshifts of the SMGs. In the bottom panel we show the residuals. The SMG redshifts agree well with our estimates obtained with the PPM for the overdensities. We found indeed an accuracy of $\sigma((z_{\rm ov} - z_{\rm spec})/(1+z_{\rm spec}))=0.043$ and a negligible bias $\langle z_{\rm ov} - z_{\rm spec} \rangle=0.034$, on average, which is competitive with the typical photometric redshift accuracy of distant galaxies \citep[e.g., in COSMOS,][]{Weaver2022} and distant protoclusters  \citep[e.g.,][]{Brinch2023}. The best accuracy is remarkably found for the overdensities around the most distant SMGs at $z\sim4-5$, for which we have $(z_{\rm ov} - z_{\rm spec})/(1+z_{\rm spec})$ between 0.019 and 0.038. These results show that the $w$PPM is thus effective in recovering well the redshifts of the overdensities, all over the entire broad redshift range covered by the SMGs. 


\subsection{Structural properties of the overdensities}\label{sec:structural_properties_PCs}
We next used the parameters measured by the $w$PPM to characterize the structural properties of the detected overdensities. As further described in Sect.~\ref{sec:PPM}, one key quantity is $\mathcal{R}_{\rm PPM}$, which is the projected size of the overdensity measured as an angular separation from the SMG. We use  $\mathcal{R}_w$ to denote the projected size of the overdensity as found by the wavelet transform. In this case, a miscentering $\theta_{\rm ov}$ between the SMG and the wavelet-based overdensity peak is determined. Another important quantity is the number of galaxies  ($N_{\rm selected}$) with projected separations of less than $\mathcal{R}_{\rm PPM}$ from the SMG and with redshifts within $(z_{\rm ov}-\overline{\Delta z}/2;z_{\rm ov}+\overline{\Delta z}/2)$, where $z_{\rm ov}$ is the estimated redshift of the protocluster and $\overline{\Delta z}=0.3$ (see Sect.~\ref{sec:PPM}). For a given photo-$z$ catalog, $N_{\rm selected}$ is thus a proxy for the richness of the detected overdensity.
In Table~\ref{tab:cluster_properties} we report these quantities for all overdensities associated with the SMGs. In the following we discuss the mutual dependence of these overdensity properties.

\subsubsection{Richness}
Overdensities detected using the B19 catalog show the highest richness, at fixed SMG, up to values of $N_{\rm selected}\simeq200$, while much lower values $N_{\rm selected}\simeq10-30$ are found for the overdensities detected with the L18 or AH18 catalogs. This is not surprising, as the B18 catalog contains $\sim(10-20)$ times more sources than both L18 and AH18 samples (see Sect.~\ref{sec:photoz_catalogs}).
Furthermore, we find that when we use the B19 catalog the overdensities associated with lower-redshift SMGs in our sample (i.e., $z<2.5$) are 60-70\% richer ($\sim$150-200) in members than those located at higher redshifts. 
This behavior is expected as indeed the photometric redshift catalog is flux-limited {and thus galaxy number densities decrease from lower to higher-redshift protoclusters.}
To some extent, the behavior can be also explained by the fact that at lower-$z$ we are looking at a significantly more
advanced stage of cluster formation, while at higher-$z$ the overdensity cores embedded in the large-scale protocluster structures {are less overdense} and in the phase of assembly.

\subsubsection{Size and redshift}
\begin{figure}[h!]
\centering
\includegraphics[trim={3.5cm 0 2.1cm 0},width=5cm]{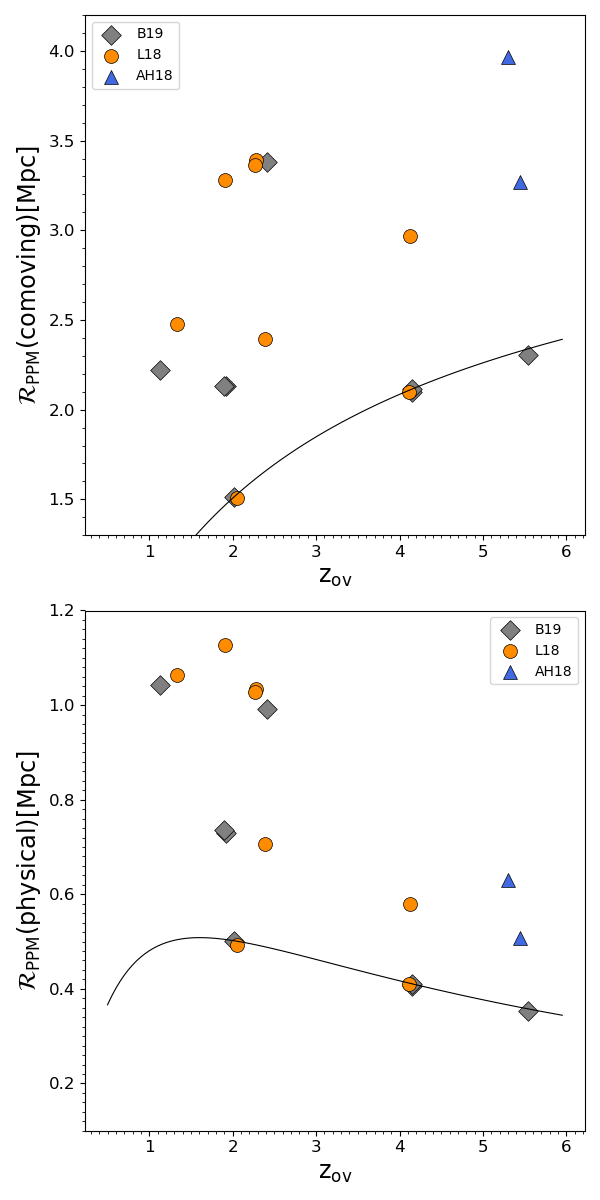}
\caption{Overdensity core sizes in comoving units (top panel) and physical units (bottom panel) as a function of overdensity redshifts estimated by the PPM. The solid black lines show the evolution with redshift of the comoving (top) and physical (bottom) core sizes that correspond to an angular aperture of 60~arcsec. Different colors and symbols refer to the different photometric redshift catalogs used to detect the overdensities, as illustrated in the legend.}
\label{fig:PCsize_vs_z}
\end{figure}

\begin{figure}[h!]
\centering
\includegraphics[trim={3.5cm 0 2.1cm 0},width=5cm]{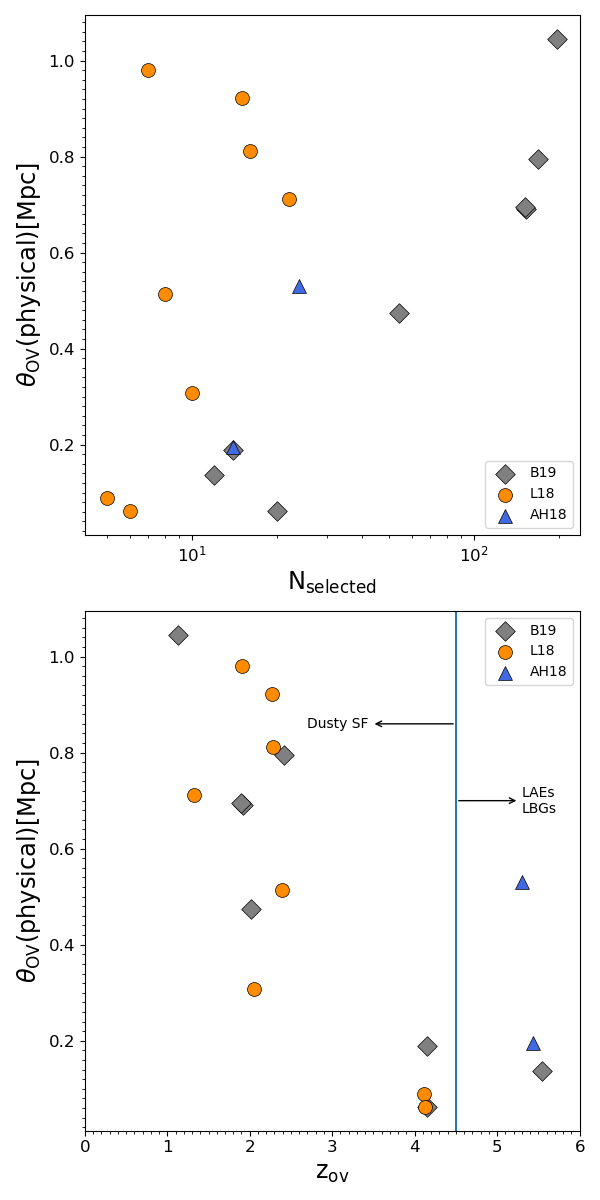}
\caption{Projected separation between the SMG coordinates and the overdensity peak as found by the wavelet transform of $w$PPM as a function of the estimated richness (top) and redshift (bottom) of the overdensity. The vertical line at $z_{\rm ov}\gtrsim4$ in the bottom panel is used to indicate the possible transitioning phase in the protocluster galaxy population. The color coding is the same as in Fig.~\ref{fig:PCsize_vs_z}.}
\label{fig:miscentering}
\end{figure}

In Figs.~\ref{fig:PCsize_vs_z} and \ref{fig:miscentering}  we compare relevant quantities estimated with the $w$PPM for the different overdensities. For each SMG, in the case of multiple overdensity associations (see Sect.~\ref{sec:multiple_associations}), we considered only the one closer to the SMG redshift. Fig.~\ref{fig:PCsize_vs_z} displays the overdensity size (in comoving and physical unit) as a function of the overdensity redshifts estimated with the PPM, for all all three photo-$z$ catalogs. We overplot as solid lines the projected radius that corresponds to the minimum size allowed by the PPM tessellation of the projected space and equal to 60~arcsec. For the sake of simplicity we report only the $\mathcal{R}_{\rm PPM}$ values. We verified indeed that our results do not change when $\mathcal{R}_{w}$ is instead plotted.  Ultimately, this is because the relative median difference between $\mathcal{R}_{\rm PPM}$ and $\mathcal{R}_{w}$ is somehow limited and equal to $\mathcal{R}_{w}/\mathcal{R}_{\rm PPM}-1 =0.15^{+0.59}_{-0.46}$, where the reported uncertainties correspond to the 68\% confidence interval. Altogether, the estimated overdensity sizes range between (1.5-4)~Mpc (comoving) as well as between (0.4-1.0)~Mpc (physical), quite independently of the redshift. We stress that these values are those associated with protocluster cores, while the total extent of large-scale protoclusters can reach (20-30)~Mpc \citep[physical,][]{Muldrew15,Casey16}. These results further highlight the capability of the PPM in detecting megaparsec-scale overdensities around the SMGs, likely associated with the cores of larger-scale protoclusters.

\subsubsection{Miscentering between the SMG and the overdensity peak.}
Figure~\ref{fig:miscentering} shows instead the projected miscentering $\theta_{\rm ov}$ between the SMG and the overdensity peak as found with the wavelet transform, as a function of both the estimated overdensity richness $N_{\rm selected}$ and the overdensity redshift $z_{\rm ov}$. Interestingly, $\theta_{\rm ov}$ has a broad range of values, from negligible offsets up to separations of $\simeq1$~Mpc (physical). Any clear trend of $\theta_{\rm ov}$ is observed neither with $N_{\rm selected}$ nor with $z_{\rm ov}$.

However, we do find evidence that the lowest $\theta_{\rm ov}\lesssim0.4$~Mpc are preferentially associated with low richness overdensities ($N_{\rm selected}\lesssim20$). The majority of the detected overdensities have instead larger  $\theta_{\rm ov}\simeq(0.4-1.0)$~Mpc, and all rich overdensities ($N_{\rm selected}\gtrsim20$) have such large $\theta_{\rm ov}$ values. These results suggest that, in the cases of large $\theta_{\rm ov}$, either the SMG is  located in the outskirts of the overdensity core or that, in the megaparsec-scale surroundings of the SMG, there exists another, more overdense region that is found with the wavelet analysis. Visual inspection of the overdensity density maps in Appendix~\ref{app:ppmplots} indeed suggests that the morphology of the large-scale protoclusters is complex. In some cases (e.g., GN10, GN20, and GN20.2a) they have compact morphologies, similarly to  CL~J1001+0220 at $z=2.51$ in the COSMOS field  \citet{Wang16}. In other cases (e.g., HDF850.1, SMM~J123634, SMM~J123606, SMM~J123618, SMM~J123707, SMM~J123711, and SMM~J123712) our detected overdensities have evidence for large-scale filamentary or clumpy structures, as typically found in still-assembling protoclusters such as SSA22 at $z=3.09$ \citep{Steidel98,Umehata2019}, the Spiderweb at $z=2.16$ \citep{Jin21}, and Hyperion at $z=2.45$ \citep{Cucciati2018}. These overdensities with nontrivial morphology represent a piece of indirect evidence for the complex dynamical state of the associated protoclusters, which are indeed non-relaxed structures.

Altogether, the measured miscentering values correspond to only a small fraction of the total extent of protoclusters, as they have overall sizes of 10-20~Mpc \citep[physical; e.g.,][]{Jin21}. Some of our SMGs could still be potentially associated with brightest cluster galaxies as indeed found in previous studies \citep[][]{Emonts2013,Zhou2020}. However, as outlined in Sect.~\ref{sec:SMG_env_general}, different SMGs may be associated with the same structure, so the identification of the proto-brightest cluster galaxies among  the SMGs in our sample is not straightforward.

\begin{figure}[h!]
\captionsetup[subfigure]{labelformat=empty}
\subfloat[]{\hspace{0.cm}\includegraphics[page=1, trim={0.7cm 0.4cm 2cm 1.8cm},clip,width=0.25\textwidth,clip=true]{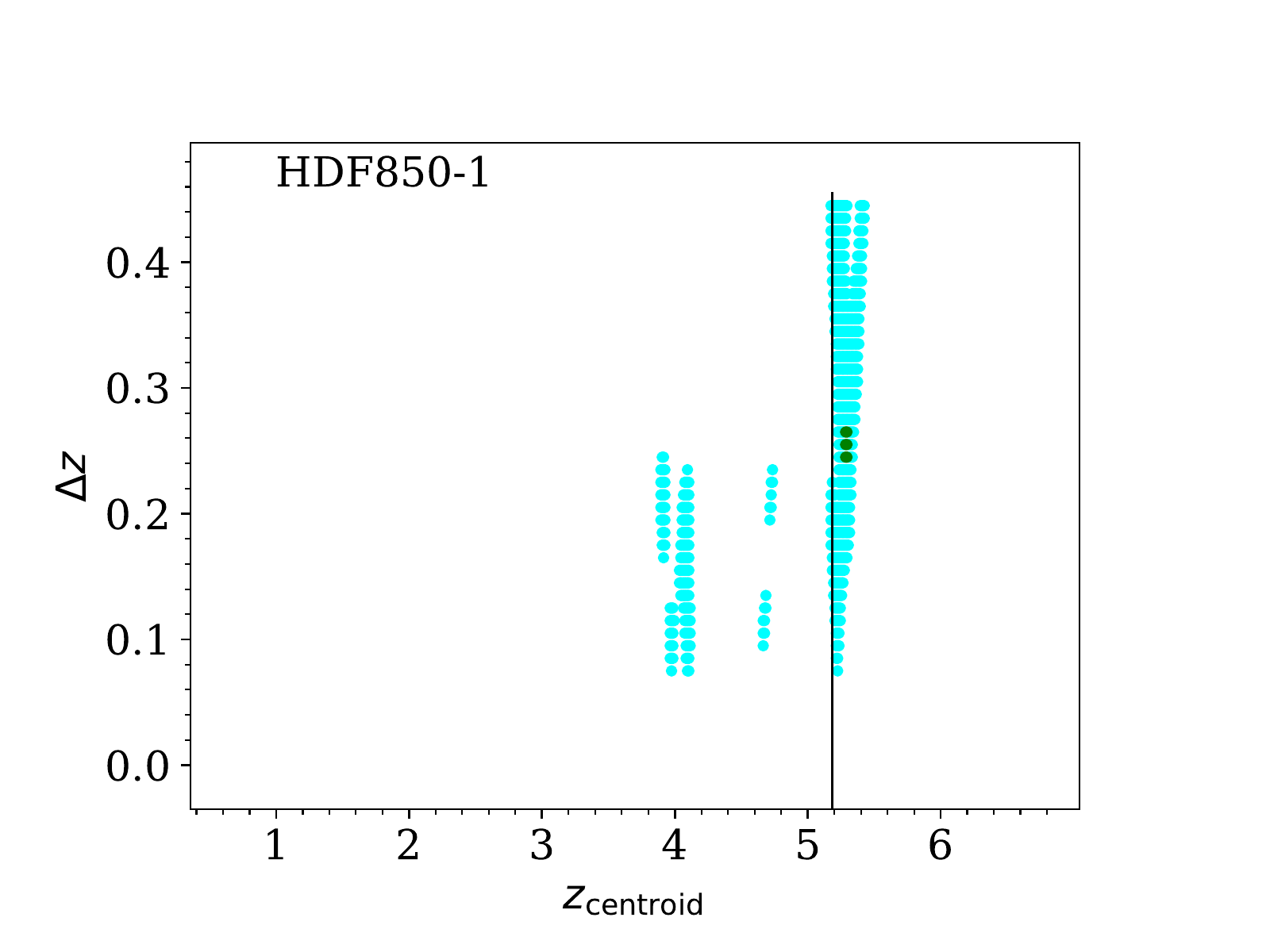}}
\subfloat[]{\vspace{0.cm}\includegraphics[page=1, trim={2cm 0.4cm 1cm 1.0cm},clip,width=0.3\textwidth,clip=true]{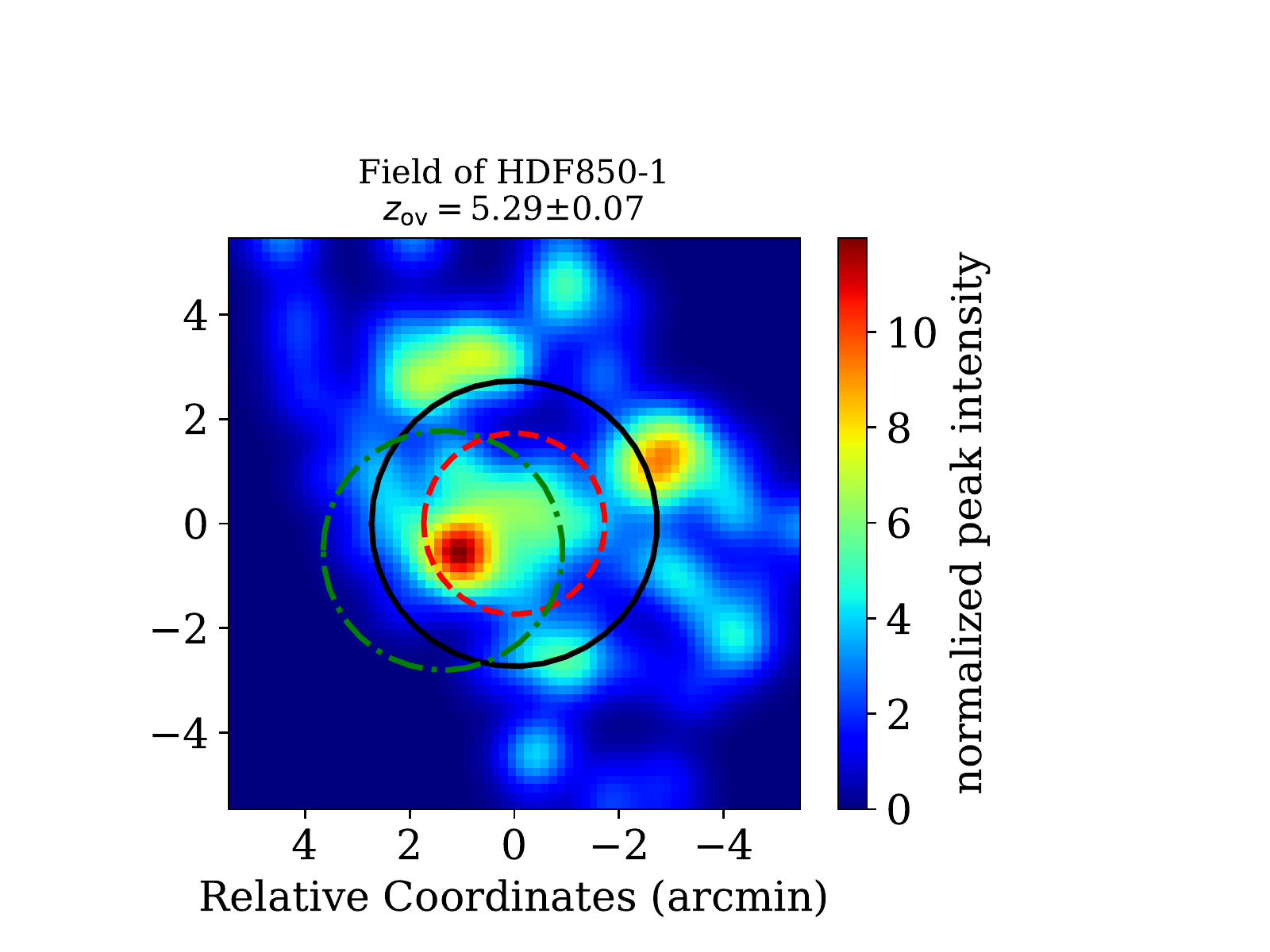}}
\vspace{-0.8cm}\\
\subfloat[]{\includegraphics[page=7, trim={0.7cm 0.4cm 2cm 1.8cm},clip,width=0.25\textwidth,clip=true]{./ppm_plots_B19.pdf}}
\subfloat[]{\hspace{0.cm}\includegraphics[page=1, trim={2cm 0.4cm 1cm 1.0cm},clip,width=0.3\textwidth,clip=true]{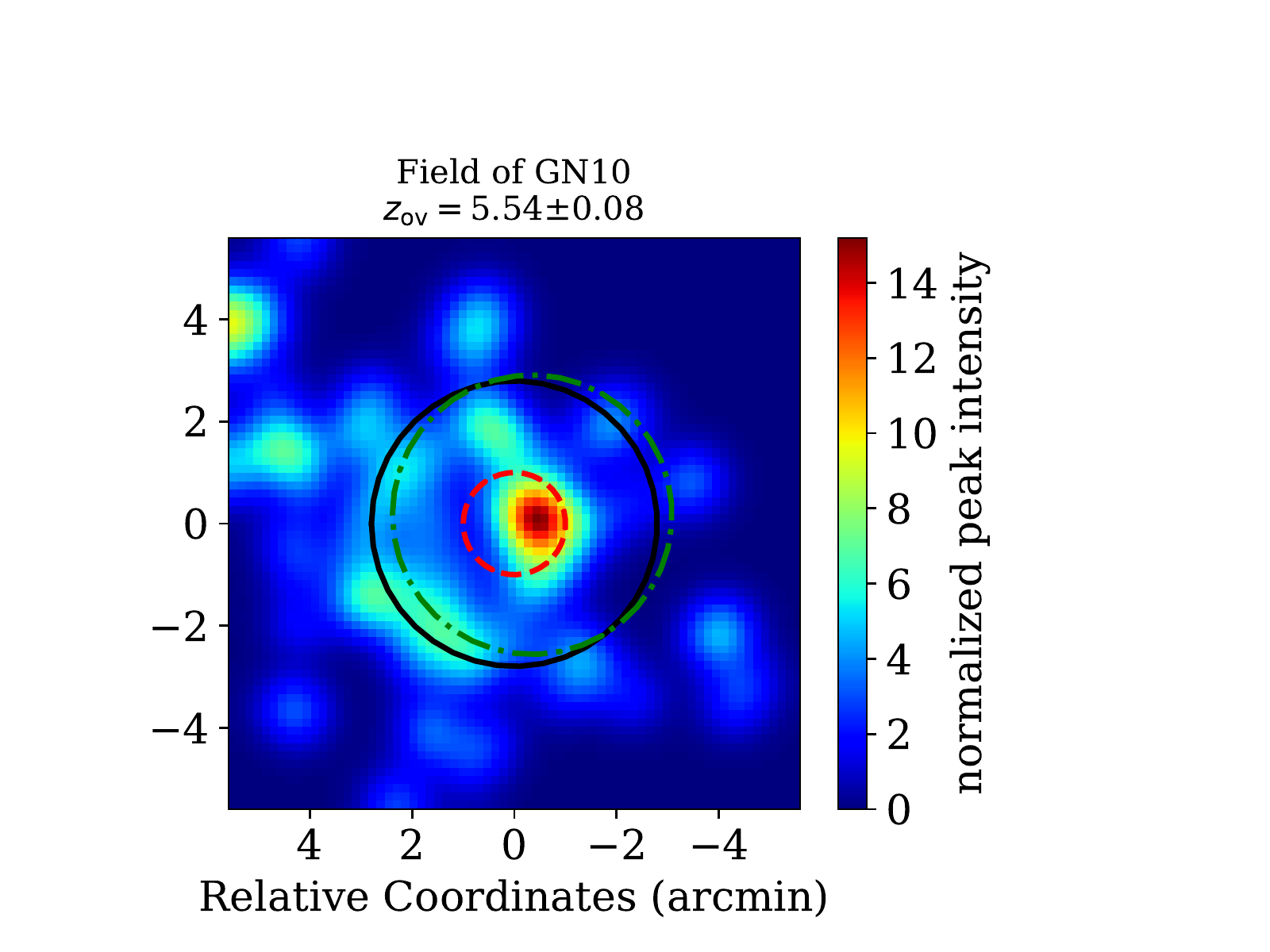}}\vspace{-0.8cm}\\
\subfloat[]{\hspace{0.cm}\includegraphics[page=8, trim={0.7cm 0.4cm 2cm 1.8cm},clip,width=0.25\textwidth,clip=true]{./ppm_plots_L18.pdf}}
\subfloat[]{\hspace{0.cm}\includegraphics[page=1, trim={2cm 0.4cm 1cm 1.0cm},clip,width=0.3\textwidth,clip=true]{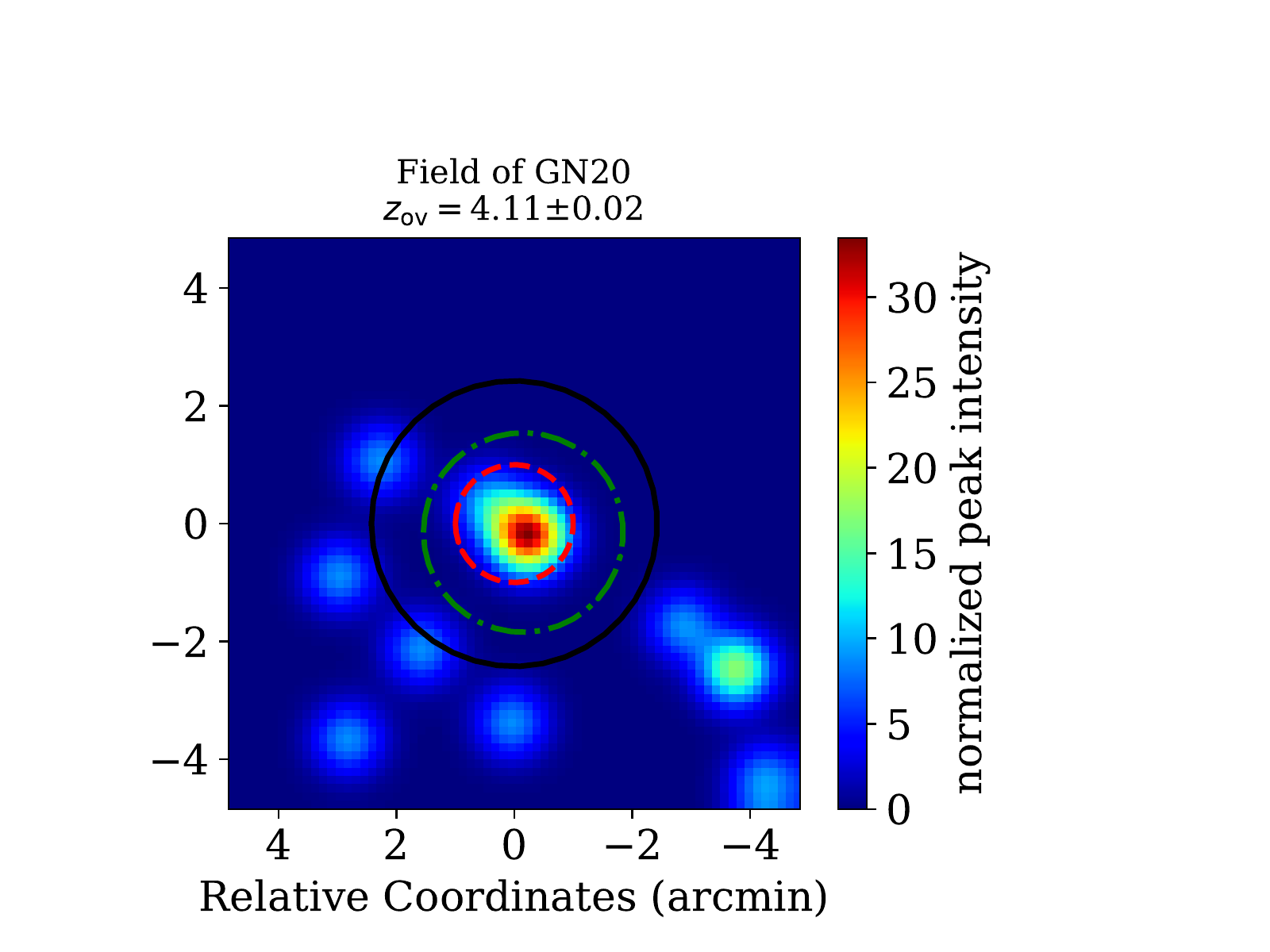}}\\
\caption{PPM plots (left) and density maps (right) of protoclusters detected around $z>4$ SMGs using the AH18 (top), B19 (center), and L18 (bottom) catalogs. {\it Left:} PPM plots for HDF850.1, GN10, and GN20. The vertical solid lines show the SMG spectroscopic redshifts. Colored dots refer to significance levels $>2\sigma$ (cyan), $3\sigma$ (green), $4\sigma$ (blue), $5\sigma$ (red), $6\sigma$ (brown), and $7\sigma$ (black). {\it Right}: Gaussian density maps for the overdensities, centered at the projected space coordinates of the SMGs. The solid black and dashed red circles are centered at the projected space coordinates of the SMGs. The former have a (physical) radius of 1 Mpc, estimated at $z_{\rm ov}$, while the latter, with a radius $\mathcal{R}_{PPM}$, denote the region within which the PPM detects the overdensity. The dotted-dashed green circle is centered at the peak of the detection as found by the wavelet transform and has a radius $\mathcal{R}_{w}$.}\label{fig:high_z_PCs}
\end{figure}

\subsection{{Protoclusters beyond $z=4$}}
\subsubsection{Megaparsec-scale overdensities and morphology}
We now discuss in more detail the overdensities around the most distant SMGs, namely the GN20--GN20.2a pair ($z=4.05$), HDF850.1 ($z=5.183$), and GN10 ($z=5.303$). As further outlined above, megaparsec-scale overdensities are detected around them at redshifts consistent with those of the SMGs. These SMGs are thus likely associated with large-scale protoclusters close to the epoch of reionization, which are excellent targets for the {\it James Webb} Space Telescope (JWST) and next generation spectrographs.

Figure~\ref{fig:high_z_PCs} displays the PPM plots (left) and density maps (right) for the overdensities around the three SMGs, when using the AH18, B19, and L18 catalogs, for HDF850.1, GN10, and GN20, respectively. The PPM plots show clear overdensity patterns at the redshift of the SMGs. Patterns at lower and even higher ($z\simeq6$) redshifts than the SMGs are also visible. These correspond to foreground and background overdensities. All three photo-$z$ catalogs used in this work are thus effective in detecting protoclusters at the highest redshifts.

The right panels of Fig.~\ref{fig:high_z_PCs} instead show the density maps centered around the SMG coordinates. As pointed out also in Sect.~\ref{sec:structural_properties_PCs}, the morphology of these high-$z$ overdensities appear heterogeneous. Several overdensity peaks are present around the SMGs. However, while GN10 and GN20 appear to be co-spatial with the overdensity with the highest significance in their field, the large-scale structure overdensity around HDF850.1 is more complex, with multiple peaks detected and a substantial miscentering of $\sim90$~arcsec ($\sim570$~kpc) between the SMG and the most prominent overdensity peak in the southeast.

Interestingly, \citet{Calvi21} performed a spectroscopic campaign of this overdensity and found a multicomponent system, similarly to what we find with the $w$PPM analysis of this work. They found that the overall structure is extended over 700~arcsec in projection (i.e., 4.4~Mpc) and presents a density enhancement located 200~arcsec (i.e., 1.3~Mpc) to the northeast of the SMG HDF850.1. This density enhancement is seen also in the density map of Fig.~\ref{fig:high_z_PCs}, while we additional probe another strong overdensity peak with a separation of 3~arcmin to the west of the SMG.

\subsubsection{A transition epoch at $z\gtrsim4$ for the protocluster galaxy population}
Interestingly, the two SMGs at the highest redshifts, {$z\sim5$} (i.e., HDF850.1 and GN10), are both found in overdensities only when the AH18 catalog is used. The other two photo-$z$ catalogs (L18, B19) are less effective in finding the associated overdensities. Using the L18 catalog, which primarily includes {\it Herschel} sources,  we do not find any of the two overdensities, while with the B19 photo-$z$ we are able to detect the overdensity associated with GN10, and that around HDF850.1 is more uncertain, as the PPM detects an overdensity at a higher redshift than the SMG. Given these results, we speculate that we possibly witness a transitioning phase in the protocluster galaxy population at {$z\gtrsim4$.} Protoclusters at $z\lesssim4$ are well detected with the B19 and L18 catalogs, which were built also using FIR data, including from {\it Herschel}. Therefore, the galaxy populations of these protocluster likely contain a number of dusty star-forming galaxies, in addition to the SMGs. On the other hand, the AH18 photo-$z$ catalog, thanks to which we detect the highest-redshift protoclusters, contains only LAEs and LBGs. 
One interpretation is that protoclusters around SMGs at $z\lesssim4$, thus around the peak of the cosmic SFR density, are rich in dusty galaxies, while $z\gtrsim4$ protoclusters are likely populated by less dusty galaxies, which appear as LAEs or LBGs (see also Fig.~\ref{fig:miscentering}, bottom panel, for a schematic classification). Nevertheless, it is necessary to exercise caution in accepting this explanation outright, and further investigation with large samples is needed to firmly confirm the proposed scenario.

These results are in agreement with those by \citet{Malavasi21}.
By studying the galaxy population of a protocluster at $z=4$ traced by LAEs, they showed that lower-redshift LAEs are on average significantly dustier than their counterparts at higher redshifts, which supports our proposed scenario for $z\gtrsim4$ being a transitioning epoch for the protocluster galaxy population.
Similarly, we note that \citet{Calvi21} recently performed a spectroscopic campaign targeting the protocluster around HDF850.1 and found that, beside HDF850.1, none of the confirmed members are bright in the FIR/(sub)millimeter wavelength range. Likewise, \citet{Daddi09}  detected the GN20--GN20.2a protocluster as an overdensity of B-band dropout LBGs. Overall, these independent results  support our proposed scenario that $z\gtrsim4$  corresponds to a transitioning epoch for the protocluster galaxy population.

\subsection{SMG molecular gas masses and the overdensities}
One of the main goals of this work is to test whether SMGs are tracers of protoclusters. Therefore, in order to investigate the possible physical interplay between the SMGs and their large-scale environments in Fig.~\ref{fig:MH2vsSNR} we plot the molecular H$_2$ gas mass of the SMGs in our sample as a function of the associated overdensity significance, as found with the PPM. In the case of multiple detections for a given SMG, we averaged the different values  obtained with the photo-$z$ catalogs used in this work, listed in Table~\ref{tab:cluster_properties}.

Remarkably, we do observe a good correlation between M$_{\rm H_2}$ and the significance of the detected overdensity. The Pearson and Spearman tests tell us that the probability that the null hypothesis (i.e., no correlation) is rejected is at a level of 3.7$\sigma$ and 2.6$\sigma$, respectively   (i.e., $p$-value$=2.36\times10^{-4}$ and $9.76\times10^{-3}$). 
{Furthermore, the highest values of $M_{\rm H_2}\simeq10^{11}~M_\odot$ and highest overdensity significance are found in the case of the highly star-forming companions GN20 and GN20.2a at $z\simeq4.05$, which have high FIR luminosities typical of HyLIRGs (see Sect.~\ref{sec:SMGprop}). This result was not predictable a priori as indeed high-$z$ overdensities are  associated with lower number counts of galaxies than those at lower redshifts and thus in principle more difficult to detect with a high significance.

Altogether, as pointed out in Sect.~\ref{sec:molecular_gas_prop}, our SMGs have, on average, the high SFRs and FIR luminosities typical of ULIRGs, which are often associated with gas-rich mergers {\citep[e.g.,][]{Sanders_Mirabel1996,Downes_Solomon1998,Genzel2001}.}
 Altogether, given the overall stellar and gas properties of the SMGs, we interpret  the  $M_{\rm H_2}$ versus overdensity significance correlation as a result of the fact that the SMGs that live in the strongest overdensities are also those that experienced most the interaction with the companions, as it may be the case for the GN20--GN20.2a pair. This interaction likely triggered gas infall and high levels of gas content and star formation in the SMGs \citep[see, e.g.,][for similar examples of distant gas rich galaxies in dense environments]{Castignani2018,Noble2019}.  Galaxy mergers are indeed often associated with ULIRGs and high levels of IR luminosities $\gtrsim10^{12}~L_\odot$ as for example in local major merging gas-rich pairs such as the Antennae Galaxies \citep{Gao2001,Ueda2012,Whitmore2014}.

\begin{figure}[h!]
\centering
\includegraphics[trim={1.5cm 0 1.1cm 0},width=0.43\textwidth]{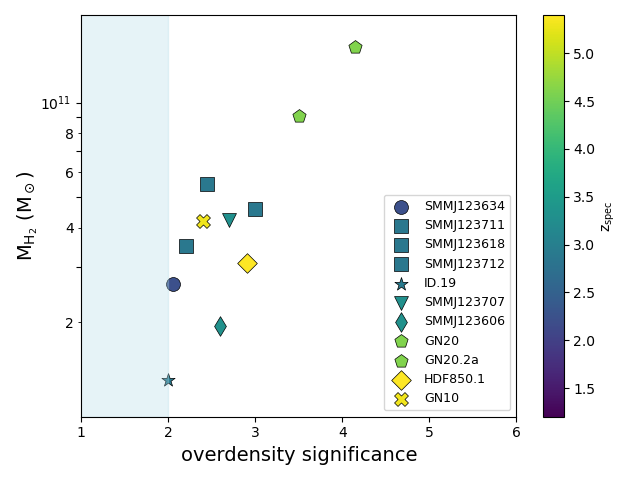}
\caption{Molecular H$_2$ gas mass for each SMG in our sample versus the average overdensity significance (in units of $\sigma$) of the associated overdensity. The points are colored according to the SMG redshift, and different symbols refer to different SMGs. {The shaded area denotes the minimum overdensity significance (= 2) that is adopted throughout this work.}}
\label{fig:MH2vsSNR}
\end{figure}

\subsection{Are SMGs good tracers of protoclusters?}

In Table~\ref{tab:targets} we present the outcome of our analysis and summarize the list of overdensities that we detected with the $w$PPM around the SMGs. Secure overdensity detections associated with the SMGs are indicated as ``yes,'' non-detections as ``no,'' and tentative detections as ``possible'' when the estimated redshift of the overdensity is only marginally consistent with that of the SMG, that is, the absolute value of the difference between the two is $\gtrsim0.3$. We also denote both GN20 and GN20.2a as ``edge'' in Col. (6) because they are located at the edge of the AH18 survey footprint. 

As outlined in Sect.~\ref{sec:SMG_env_general}, some SMGs do belong to the same known overdensities as they are close to each other both in redshift and in projected coordinates. These are the triplet SMM~J123711, SMM~J123618, and SMM~J123712 as well as the GN20--GN20.2a pair. For these sources, we obtain similar results when looking for overdensities in the three photo-$z$ catalogs used in this work. The only exception is SMM~J123618, for which we find an overdensity only in the L18 catalog, at variance with SMM~J123711 and SMM~J123712, for which overdensities are found both in B19 and L18.

In the case of the SMG ID.03, an overdensity is detected both in the B19 and AH18 photometric redshift catalogs but at a lower redshift; thus, we classified these overdensities as possible.  Interestingly, the source ID.03 is the only SMG in our sample that is not associated with an overdensity in at least one of the three photometric redshift catalogs. ID.03, as well as ID.19 for which we detect a relatively poor $2\sigma$ overdensity, have the lowest gas masses $M_{\rm H_2}\simeq(1-2)\times10^{10}~M_\odot$  among those reported in Table~\ref{tab:mol_gas_properties} for our SMG sample. These relatively low H$_2$ gas masses imply that the two sources may not be true SMGs, that is, with strong dust-continuum emission and high molecular gas content. Furthermore, the two sources are not clearly detected in several CO(J$\rightarrow$J-1) lines, as it is often the case for SMGs. Indeed, ID.19 has been detected in CO(3$\rightarrow$2) only \citep{Decarli14}, while ID.03 has relatively low-J detections by \citet{Decarli14} and \citet{Boogaard2023} in CO(1$\rightarrow$0) (tentative), CO(2$\rightarrow$1), and CO(4$\rightarrow$3).

Altogether, for 11 out of 12 SMGs we detect an overdensity in at least one of the three photometric catalogs. This corresponds a success rate of $92\%\pm8\%$, where we  report the average value as well as the root mean square uncertainty derived using the binomial distribution \citep[see, e.g.,][]{Castignani2014b}. When grouping the above mentioned SMGs that likely belong to the same parent protoclusters we find eight overdensities, while we recover all three previously known protoclusters. We thus more than double the number of known overdensities physically associated with SMGs in the GOODS-N field. 

Interestingly, while we find a high fraction (92\%) of SMGs in protoclusters, with a similar PPM-based search, \citet{Castignani2014b} found a lower fraction of $\sim70\%$, when considering instead distant radio galaxies at $z\sim1-2$ in the COSMOS field, which suggests that bright SMGs may be better tracers of protocluster cores. We stress, however, that this comparison relies on different data sets and should be taken with caution, considering in particular the advent of forthcoming low-frequency radio surveys, for example with the Square Kilometer Array, which will enable the detection of distant radio-loud active galactic nuclei and radio galaxies down to the lowest radio powers. 

In Table~\ref{tab:targets} (column 7) we report the overdensity success rate over all three considered photo-$z$ catalogs. We find that the vast majority of the SMGs are found in overdensities with the B19 (8 out of 12 SMGs), L18 (9 out of 12 SMGs), and AH18 (2 out of 3 SMGs) catalogs, separately, where we took under consideration the fact that the AH18 catalog does not have sources at $z<3.35$. These results correspond to similar success rates of 67\%, 75\%, and 67\%  for the three photo-$z$ catalogs, respectively.

We stress here that PPM overdensities are detected down to $2\sigma$ significance, similarly to our previous studies \citep{Castignani2014a,Castignani2014b,Castignani2019}. This is not unusual as indeed distant megaparsec-scale overdensities within larger-scale protoclusters are often characterized by low number counts and thus possibly detected with relatively low significance by the PPM. A notable example is the overdensity around COSMOS-FRI~03, which was detected by the PPM at $z_{\rm ov}=2.39\pm0.09$ with a $2.5\sigma$ significance \citep{Castignani2014b}, and later spectroscopically confirmed at $z=2.45$ \citep{Diener2015}. Another case is the $2.8\sigma$ PPM overdensity at $z_{\rm ov}=2.65\pm0.04$ that \citet{Castignani2019} found around the spectroscopically confirmed COSMOS-FRI~70 radio galaxy ($z=2.625$). Interestingly, if we limit ourselves to a higher significance threshold of $\geq2.5\sigma$ and $\geq3\sigma$, we find only a slightly lower number of overdensities, six and four, respectively, compared to the eight recovered using the $2\sigma$ limit of this work; the three previously confirmed protoclusters in our sample \citep{Chapman09,Daddi09,Walter12} are all detected at $\geq3\sigma$ by the PPM with a least one photo-$z$ catalog.
Altogether, we expect the PPM overdensities reported in this work to have total masses greater than $\sim10^{13}~M_\odot$, which is what \citet[][see their Sect.~8.7.1]{Castignani2014b} found by cross-matching the PPM overdensities with catalogs of distant groups in COSMOS with available mass estimates.

\begin{table}
\small\addtolength{\tabcolsep}{-4pt}
\centering
Overdensities around GOODS-N SMGs.\\
        \begin{tabular}{c|ccccccccc}
    \hline
    \hline
Name & $z_{\rm spec}$ &known &  B19 & L18 & AH18 & Success Rate  \\
(1) & (2) & (3) & (4) & (5) & (6) & (7) \\
\hline
\hline
SMM~J123634&1.225&&yes&yes &--& 2/2 \\
\hline
ID.03 &1.784&&possible& possible &--& 0/2 \\
\hline
SMM~J123711 &1.995& yes & yes&yes&--& 2/2\\
SMM~J123618&1.996&yes& no&yes&--&1/2 \\
SMM~J123712&1.996&yes&yes&yes&--&2/2\\
\hline
ID.19&2.047&&yes &yes &--&2/2\\
\hline
SMM~J123707 &2.487&&yes&yes&-- & 2/2 \\
\hline
SMM~J123606&2.505&&no &yes &--& 1/2\\
\hline
GN20&4.055& yes & yes & yes &edge & 2/2\\
GN20.2a&4.051&yes&  yes & yes &edge & 2/2 \\
\hline
HDF850.1  &5.183&yes&possible &no & yes&1/3 \\
\hline
GN10&5.303&& yes &no &yes& 2/3\\
\hline
\hline
\end{tabular}
\caption{Column description: (1) source ID; (2) CO-based spectroscopic redshift; (3) known overdensity flag; (4-6) PPM detection flags using the three photo-$z$ catalogs used in this work (yes$=$secure detection; possible$=$detection with an uncertain association of the overdensity with the SMG; no$=$no detection with a significance $\geq2\sigma$); (7) overdensity success rate among the photo-$z$ catalogs used. {\bf Note:} The dashed line in column (6) indicates that the source is outside the redshift range of the AH18 photo-$z$ catalog, while {\it edge} denotes that the source is located at the edge of the AH18 survey footprint.}
\label{tab:targets}
\end{table}

\section{Summary and conclusions}\label{sec:conclusions}
In this paper we have employed the PPM along with its wavelet-based extension \citep[$w$PPM;][]{Castignani2014a,Castignani2019} to carry out a detailed statistical analysis of the megaparsec-scale environments of distant SMGs. More specifically, we applied the  $w$PPM to search for protoclusters around a sample of 12 spectroscopically confirmed SMGs at $z\simeq1.2-5.3$ in the GOODS-N field using three photometric redshift catalogs independently. The main results of this analysis can be summarized as follows: 

\begin{itemize}
\item Of the 12 SMGs in our sample, 11 (i.e., $92\%\pm8\%$) are physically associated with megaparsec-scale overdensities. Altogether, these ten SMGs belong to eight overdensities. Three of them are previously known protoclusters, and we detect five new protoclusters between $z\sim1.2$ and $5.3$. We have thus doubled the number of known overdensities physically related to SMGs in the GOODS-N field. The detection rates for each of the three photometric catalogs range between 67\% and 75\%. 
By comparison with the spectroscopic redshifts ($z_{\rm spec}$) of the SMGs, we find that the redshifts of the detected overdensities ($z_{\rm ov}$) are well recovered, with an accuracy of $\sigma((z_{\rm ov}-z_{\rm spec})/(1+z_{\rm spec}))=0.043$. \\

\item  A wavelet-based analysis of the protocluster fields shows that the majority of the detected overdensities have separations of $(0.4-1.0)$~Mpc between the SMG and  the overdensity peak. 
Therefore, the SMGs do not always reside in the most overdense peak of the protocluster. They live in protocluster cores or overdensities that have an average size of $\sim(0.4-1)$~Mpc and are characterized by a complex morphology (compact, filamentary, or clumpy). \\

\item We find a good correlation between the {molecular (H$_2$)} gas mass of the SMGs and the overdensity detection significance. We interpret this correlation as a result of the fact that the SMGs that live in the strongest overdensities are also those that experienced the most interaction with their companions. This interaction likely triggered gas infall and high levels of gas content and star formation in the SMGs. \\

\item We speculate that we are possibly witnessing a transitioning phase at $z\simeq4$ for the galaxy population of protoclusters. While $z\lesssim4$ protoclusters appear to be mostly populated by dusty galaxies, those at higher $z$ are mostly detected as overdensities of LAEs or LBGs. However, it is essential to approach this interpretation with caution. Further studies on larger samples are needed to validate and expand upon these findings and would ultimately provide a more solid foundation for drawing definitive conclusions.

\end{itemize}

The analysis and results presented in this paper support the scenario that SMGs are excellent signposts of protoclusters. We stress that this result was made possible thanks to the use of (i) spectroscopically confirmed SMGs, (ii) good photometric redshifts for the sources in the field of the SMGs themselves, and (iii) a specific method, that is, the $w$PPM, which is tailored to find overdensities around specific tracers in the regime of low galaxy number counts and shot noise.  This is at variance with previous studies, {which used mostly photometrically selected SMGs,} that did not find a high fraction of SMGs in overdensities \citep{AlvarezCrespo2021,Gao2022}

{Altogether, the protoclusters presented in this paper are excellent targets for ongoing facilities such as JWST. Furthermore, the advent of infrared facilities such as JWST \citep[][]{Laporte2022,Morishita2022}
 and {\it Euclid} \citep{Scaramella2022}, as well as next-generation wide-field spectrographs such as MOONS \citep{Cirasuolo2020}, will provide an unprecedented leverage to constrain galaxy evolution in assembling protoclusters, which makes the present work a benchmark for future studies with larger samples of SMGs and protoclusters than those considered in this work.}

\begin{acknowledgements}
RC acknowledges partial financial support through the grant PRIN-MIUR 2017WSCC3. GC acknowledges the support from the grant ASI n.2018-23-HH.0.
RC and HD acknowledges financial support from the Agencia Estatal de Investigación del Ministerio de Ciencia e Innovación (AEI-MCINN) under grant (La evolución de los cíumulos de galaxias desde el amanecer hasta el mediodía cósmico) with reference (PID2019-105776GB-I00/DOI:10.13039/501100011033). HD acknowledges support from the ACIISI, Consejería de Economía, Conocimiento y Empleo del Gobierno de Canarias and the European Regional Development Fund (ERDF) under grant with reference PROID2020010107.   

\end{acknowledgements}

\bibliographystyle{aa}
\bibliography{paper.bib}

\onecolumn
\begin{appendix} 
\section{PPM results}\label{app:ppmplots}
In Figs.~\ref{fig:barro}, \ref{fig:liu}, and \ref{fig:arrabal} we report the PPM plots and density maps for all overdensities reported in Table~\ref{tab:cluster_properties} when using the B19, L18, and AH18 catalogs, respectively. We refer to Sect.~\ref{sec:PPM} and our previous studies \citep{Castignani2014a,Castignani2014b,Castignani2019} for details about these $w$PPM outputs.

The left panels show the PPM plots where the overdensity patterns at different redshifts along the line of sight of each SMG are displayed. In each plot the vertical solid line shows the SMG spectroscopic redshift. Colored dots refer to significance levels $>2\sigma$ (cyan), 3$\sigma$ (green), 4$\sigma$ (blue), 5$\sigma$ (red), 6$\sigma$ (brown), and 7$\sigma$ (black). 

The right panels display the Gaussian density maps centered at the projected coordinates of the SMGs. The pixel size is 1/16~Mpc while the Gaussian kernel has $\sigma=3/16$~Mpc. Sources with photometric redshifts between $z_{ov}-\overline{\Delta z}/2$ and $z_{ov}+\overline{\Delta z}/2$ were considered to produce the maps. The values of $z_{\rm ov}$ are reported in Table~\ref{tab:cluster_properties}, while $\overline{\Delta z}=0.3$ except for ID.19 (for B19 and L18 catalogs) and GN20.2a (B19 catalog), for which we chose $\overline{\Delta z}=0.2$ (see Table~\ref{tab:cluster_properties}).
The solid black and dashed red circles are centered at the projected coordinates of the SMG. The former have a (physical) radius of 1~Mpc, estimated at $z_{ov}$,  while the latter, with a radius $\mathcal{R}_{\rm PPM}$, denote the region within which the PPM detects the overdensity. The dotted-dashed green circle is centered at the peak of the detection as found by the wavelet transform and has a radius $\mathcal{R}_{w}$.

\begin{figure*}[]\centering
\captionsetup[subfigure]{labelformat=empty}
\centering
\begin{tabular}{@{}c@{}}
\subfloat[]{\hspace{1cm}\includegraphics[page=14, trim={0.0cm 0.4cm 0cm 1.0cm}, clip,width=0.45\textwidth,clip=true]{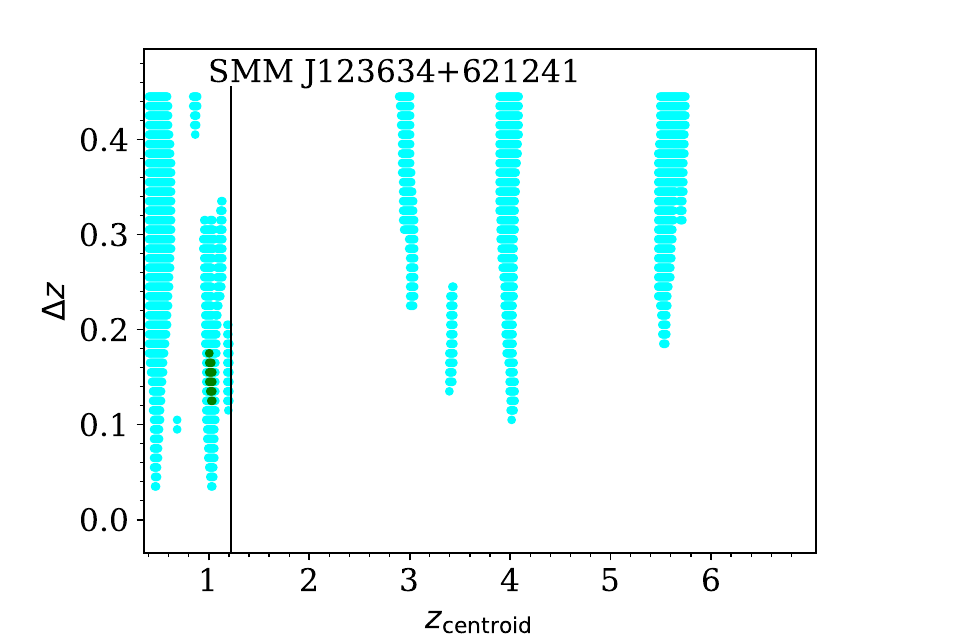}}
\end{tabular}
\begin{tabular}{@{}c@{}}
\subfloat[]{\hspace{-1.cm}\includegraphics[page=3, trim={0.0cm 0.4cm 0cm 1.0cm},clip,width=0.45\textwidth,clip=true]{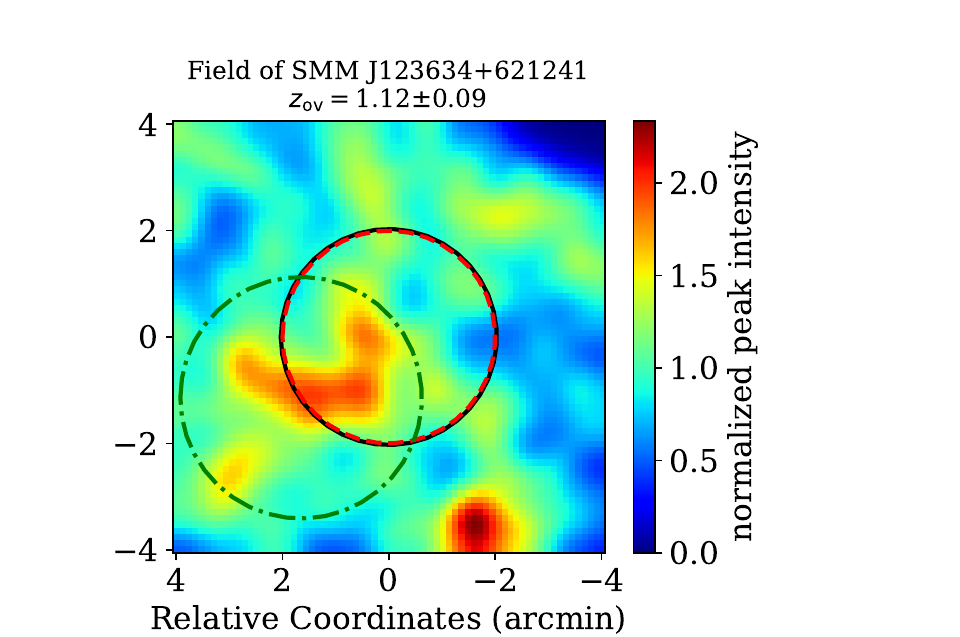}}
\\
\subfloat[]{\hspace{-1.cm}\includegraphics[page=3, trim={0.0cm 0.4cm 0cm 1.0cm},clip,width=0.45\textwidth,clip=true]{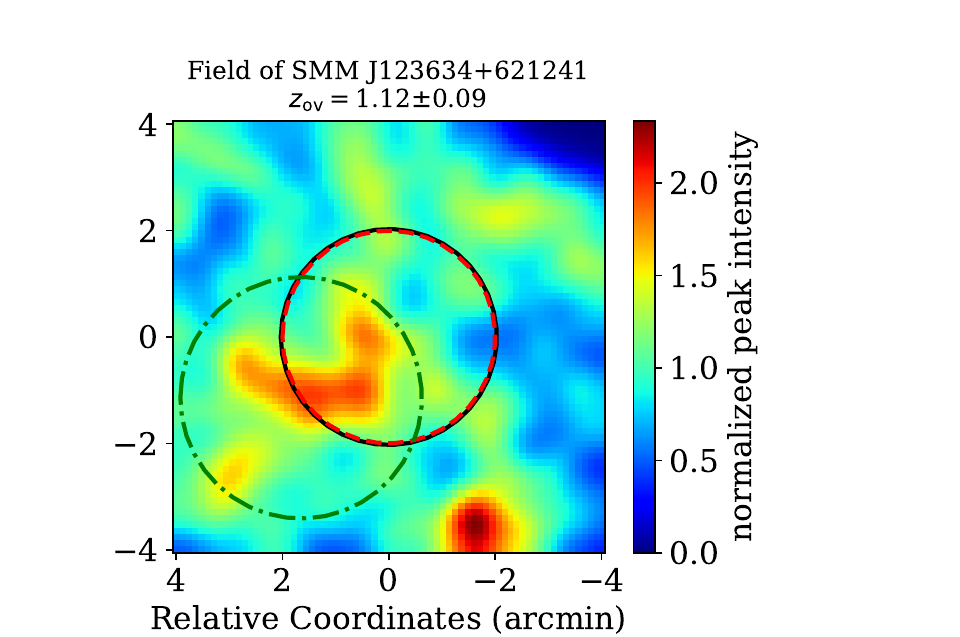}}
\end{tabular}

\vspace{-0.5cm}
\subfloat[]{\hspace{1.cm}\includegraphics[page=5, trim={0.0cm 0.4cm 2cm 1.8cm},clip,width=0.4\textwidth,clip=true]{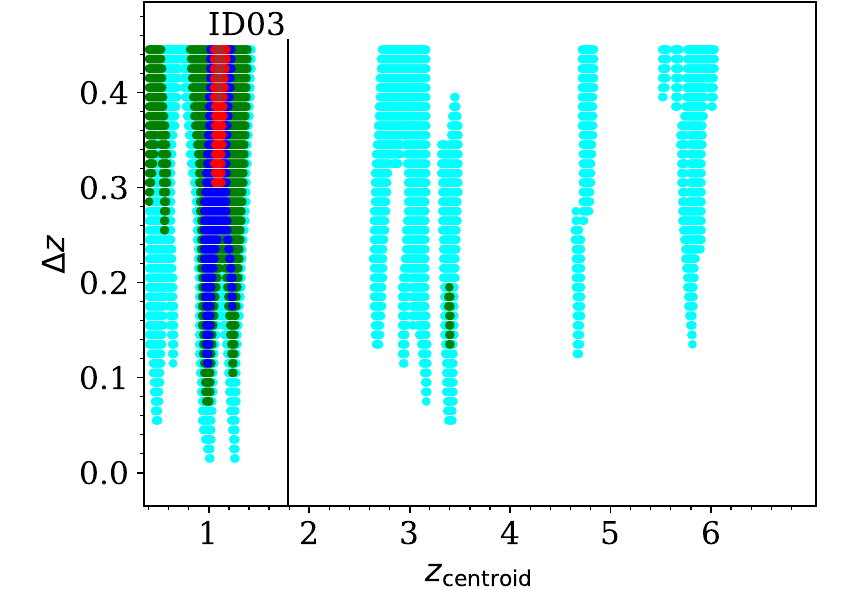}}
\subfloat[]{\hspace{0.cm}\includegraphics[page=2, trim={0.0cm 0.4cm 0cm 1.0cm},clip,width=0.45\textwidth,clip=true]{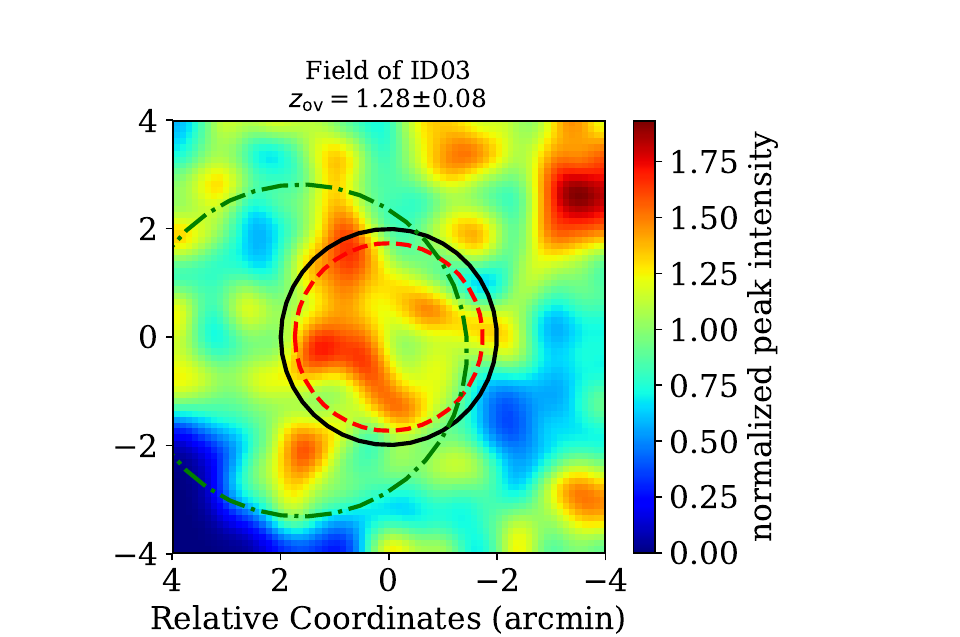}}\\
\vspace{-0.5cm}
\subfloat[]{\hspace{1.cm}\includegraphics[page=16, trim={0.0cm 0.4cm 2cm 1.8cm},clip,width=0.4\textwidth,clip=true]{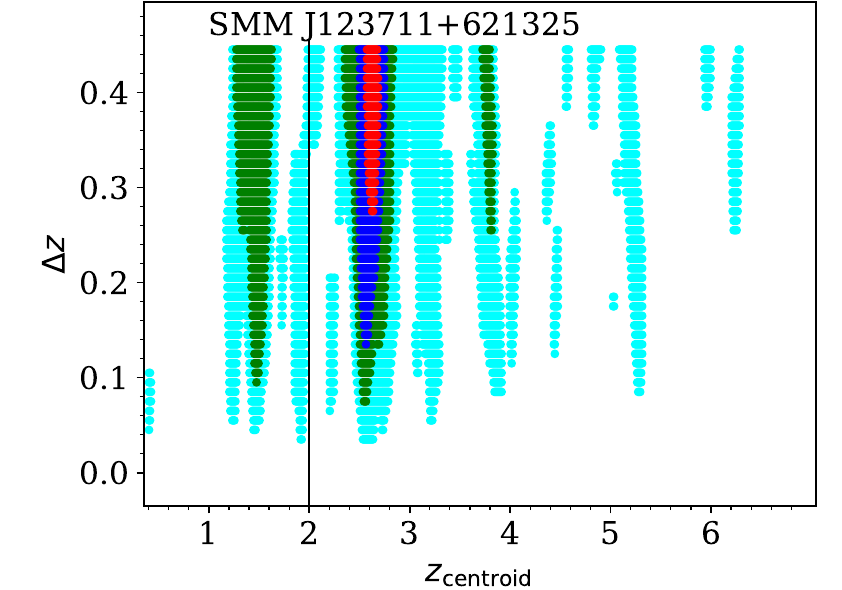}}
\subfloat[]{\hspace{0.cm}\includegraphics[page=7, trim={0.0cm 0.4cm 0cm 1.0cm},clip,width=0.45\textwidth,clip=true]{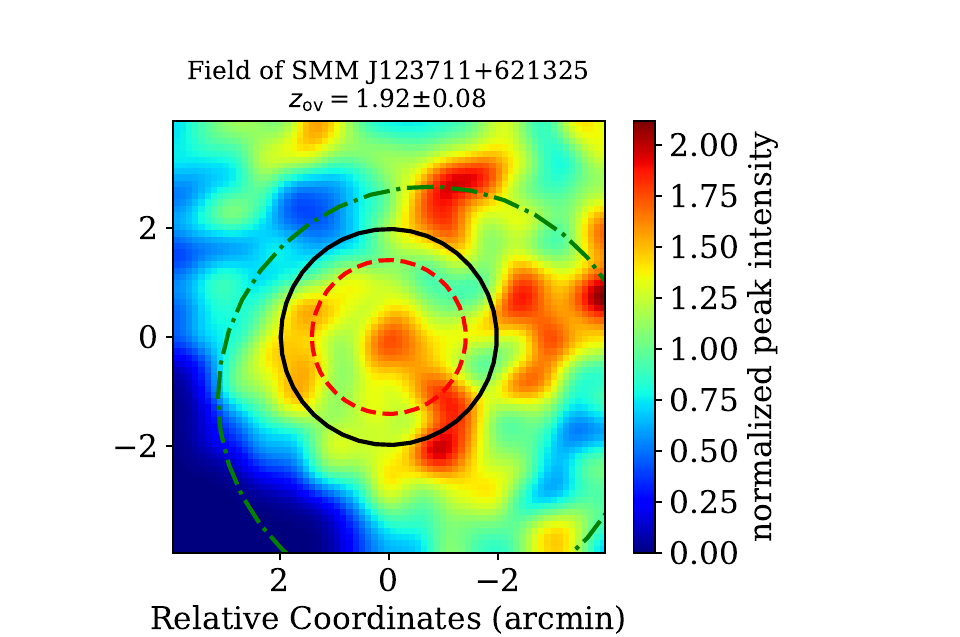}}\\
\caption{PPM plots (left) and density maps (right) for the SMGs and corresponding overdensities found using the B19 photo-$z$ catalog. See the text for details about the color coding.}
\label{fig:barro}
\end{figure*}

\begin{figure*}[]\centering
\captionsetup[subfigure]{labelformat=empty}
\ContinuedFloat
\subfloat[]{\hspace{1.cm}\includegraphics[page=17, trim={0.0cm 0.4cm 2cm 1.8cm},clip,width=0.4\textwidth,clip=true]{./ppm_plots_B19.pdf}}
\subfloat[]{\hspace{0.cm}\includegraphics[page=8, trim={0.0cm 0.4cm 0cm 1.0cm},clip,width=0.45\textwidth,clip=true]{./density_maps_Deltaz_0_295_B19.pdf}}
\vspace{-0.5cm}
\subfloat[]{\hspace{1.cm}\includegraphics[page=6, trim={0.0cm 0.4cm 2cm 1.8cm},clip,width=0.4\textwidth,clip=true]{./ppm_plots_B19.pdf}}
\subfloat[]{\hspace{0.cm}\includegraphics[page=1, trim={0.0cm 0.4cm 0cm 1.0cm},clip,width=0.45\textwidth,clip=true]{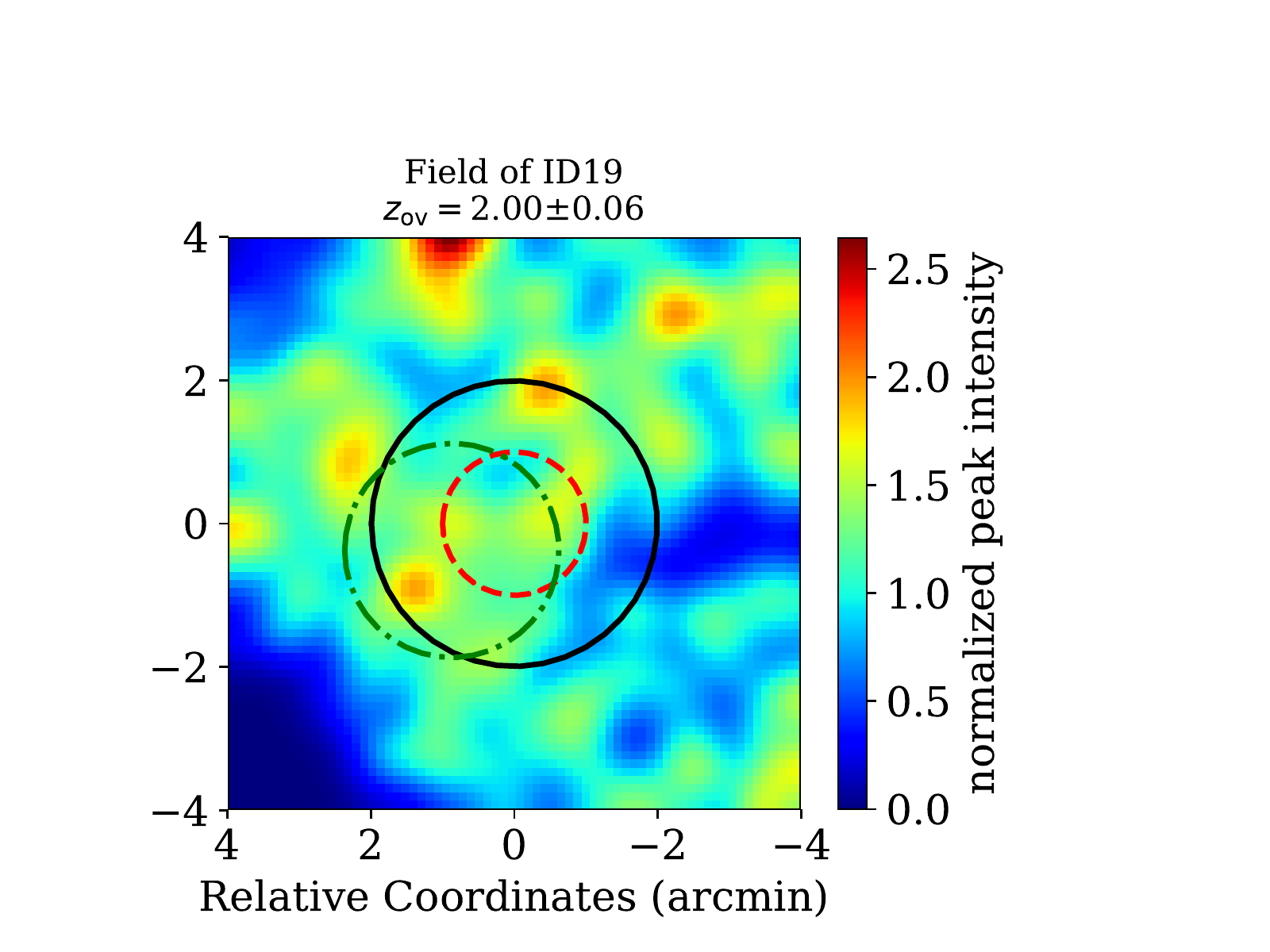}}\\

\centering
\begin{tabular}{@{}c@{}}
\subfloat[]{\hspace{1cm}\includegraphics[page=15, trim={0.0cm 0.4cm 0cm 1.0cm}, clip,width=0.45\textwidth,clip=true]{./ppm_plots_B19.pdf}}
\end{tabular}
\begin{tabular}{@{}c@{}}
\subfloat[]{\hspace{-1.cm}\includegraphics[page=5, trim={0.0cm 0.4cm 0cm 1.0cm},clip,width=0.45\textwidth,clip=true]{./density_maps_Deltaz_0_295_B19.pdf}}
\\
\subfloat[]{\hspace{-1.cm}\includegraphics[page=6, trim={0.0cm 0.4cm 0cm 1.0cm},clip,width=0.45\textwidth,clip=true]{./density_maps_Deltaz_0_295_B19.pdf}}
\end{tabular}

\caption{Continued.}
\label{fig:barro}
\end{figure*}

\begin{figure*}[]\centering
\captionsetup[subfigure]{labelformat=empty}
\ContinuedFloat
\subfloat[]{\hspace{1.cm}\includegraphics[page=8, trim={0.0cm 0.4cm 2cm 1.8cm},clip,width=0.4\textwidth,clip=true]{./ppm_plots_B19.pdf}}
\subfloat[]{\hspace{0.cm}\includegraphics[page=2, trim={0.0cm 0.4cm 0cm 1.0cm},clip,width=0.45\textwidth,clip=true]{./density_maps_Deltaz_0_295_B19.pdf}}\\
\vspace{-0.5cm}
\subfloat[]{\hspace{1.cm}\includegraphics[page=9, trim={0.0cm 0.4cm 2cm 1.8cm},clip,width=0.4\textwidth,clip=true]{./ppm_plots_B19.pdf}}
\subfloat[]{\hspace{0.cm}\includegraphics[page=4, trim={0.0cm 0.4cm 0cm 1.0cm},clip,width=0.45\textwidth,clip=true]{./density_maps_Deltaz_0_200_B19.pdf}}\\
\vspace{-0.5cm}
\subfloat[]{\hspace{1.cm}\includegraphics[page=1, trim={0.0cm 0.4cm 2cm 1.8cm},clip,width=0.4\textwidth,clip=true]{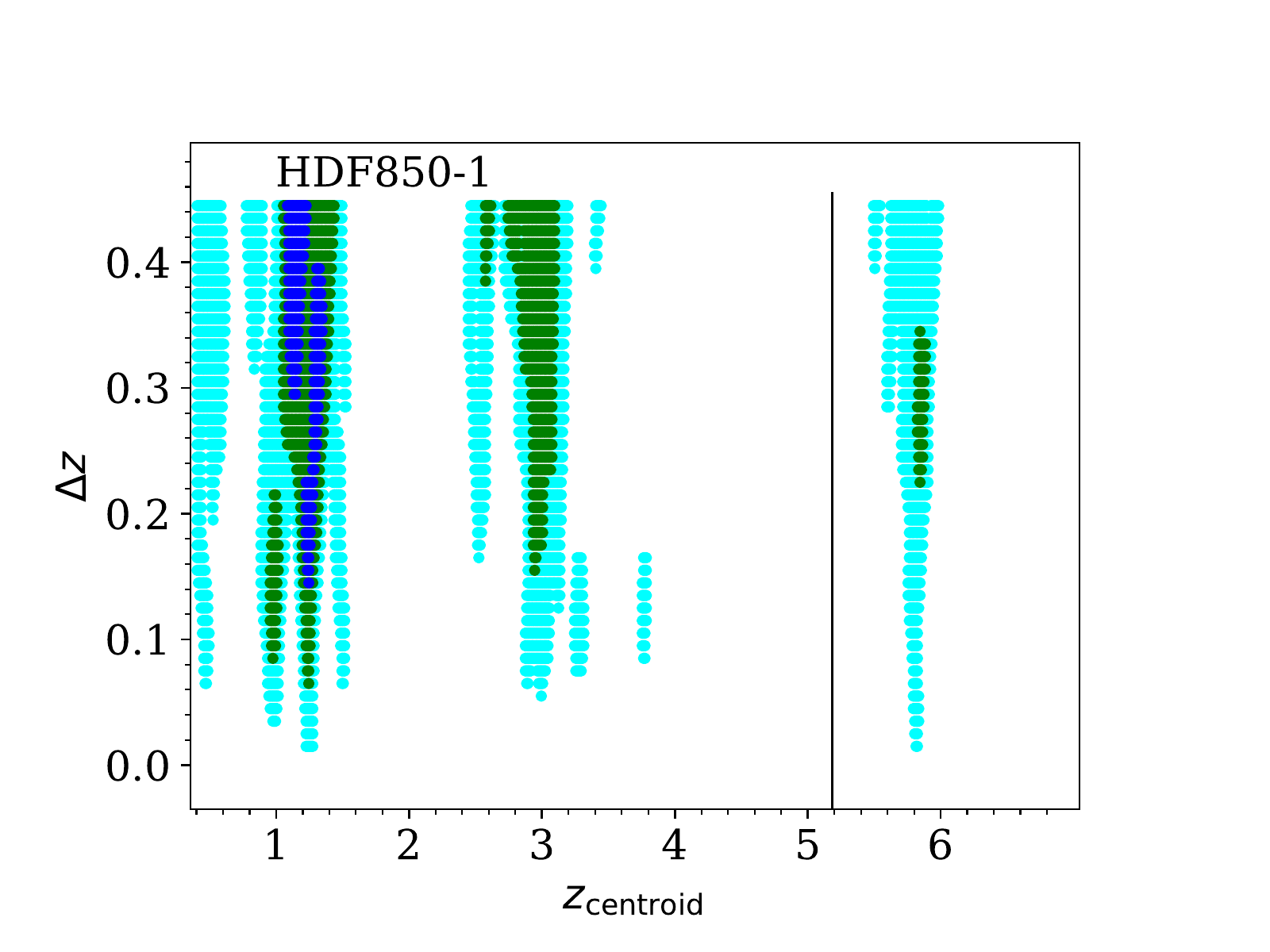}}
\subfloat[]{\hspace{0.cm}\includegraphics[page=1, trim={0.0cm 0.4cm 0cm 1.0cm},clip,width=0.45\textwidth,clip=true]{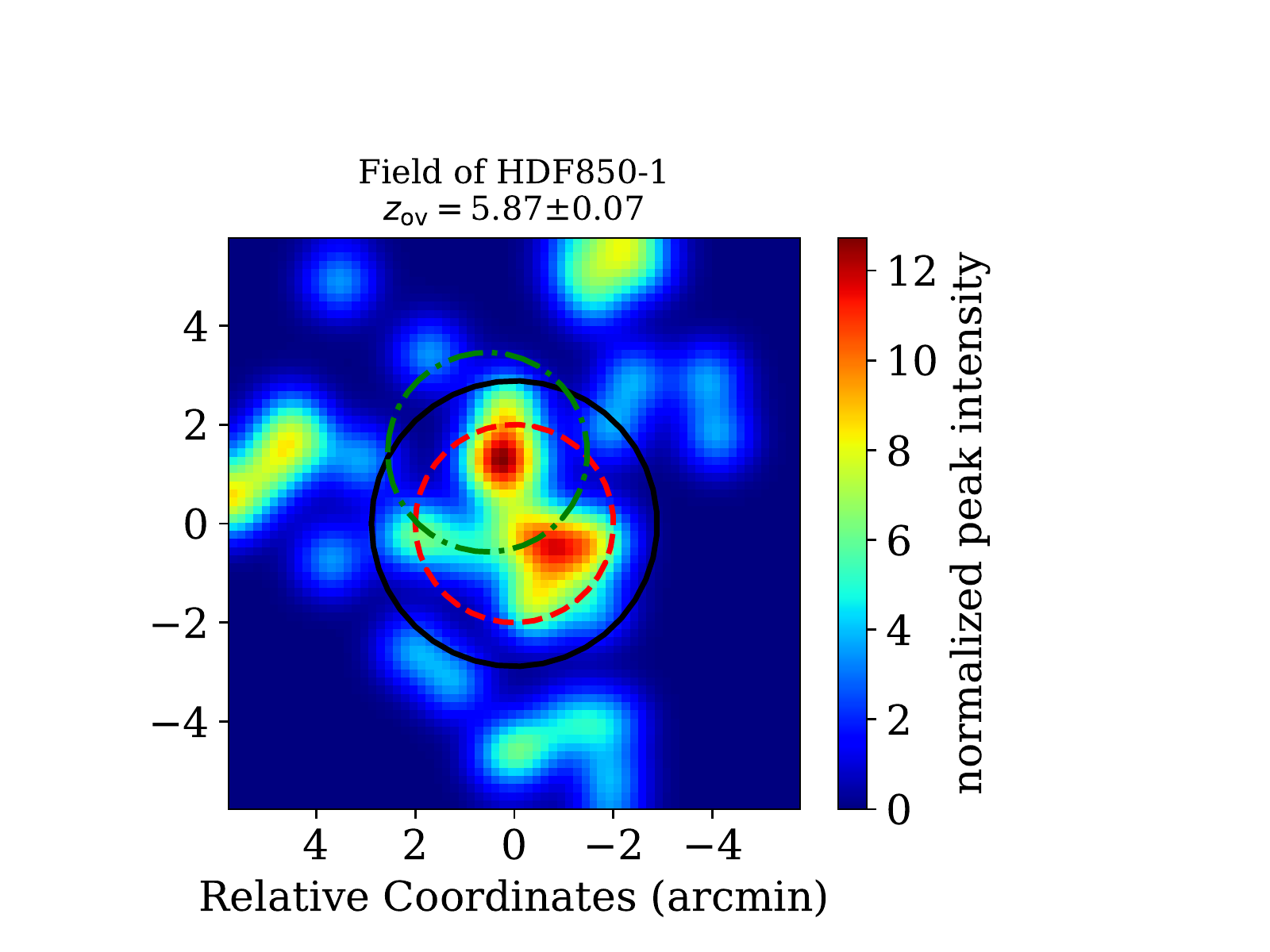}}\\
\vspace{-0.5cm}
\subfloat[]{\hspace{1.cm}\includegraphics[page=7, trim={0.0cm 0.4cm 2cm 1.8cm},clip,width=0.4\textwidth,clip=true]{./ppm_plots_B19.pdf}}
\subfloat[]{\hspace{0.cm}\includegraphics[page=1, trim={0.0cm 0.4cm 0cm 1.0cm},clip,width=0.45\textwidth,clip=true]{./density_maps_Deltaz_0_295_B19.pdf}}\\
\caption{Continued.}
\label{fig:barro}
\end{figure*}

\begin{figure*}[]\centering
\captionsetup[subfigure]{labelformat=empty}
\centering
\begin{tabular}{@{}c@{}}
\subfloat[]{\hspace{1cm}\includegraphics[page=14, trim={0.0cm 0.4cm 0cm 1.0cm}, clip,width=0.45\textwidth,clip=true]{./ppm_plots_L18.pdf}}
\end{tabular}
\begin{tabular}{@{}c@{}}
\subfloat[]{\hspace{-1.cm}\includegraphics[page=5, trim={0.0cm 0.4cm 0cm 1.0cm},clip,width=0.45\textwidth,clip=true]{./density_maps_Deltaz_0_295_L18.pdf}}
\\
\subfloat[]{\hspace{-1.cm}\includegraphics[page=6, trim={0.0cm 0.4cm 0cm 1.0cm},clip,width=0.45\textwidth,clip=true]{./density_maps_Deltaz_0_295_L18.pdf}}
\end{tabular}

\vspace{-0.5cm}
\subfloat[]{\hspace{1.cm}\includegraphics[page=5, trim={0.0cm 0.4cm 2cm 1.8cm},clip,width=0.4\textwidth,clip=true]{./ppm_plots_L18.pdf}}
\subfloat[]{\hspace{0.cm}\includegraphics[page=1, trim={0.0cm 0.4cm 0cm 1.0cm},clip,width=0.45\textwidth,clip=true]{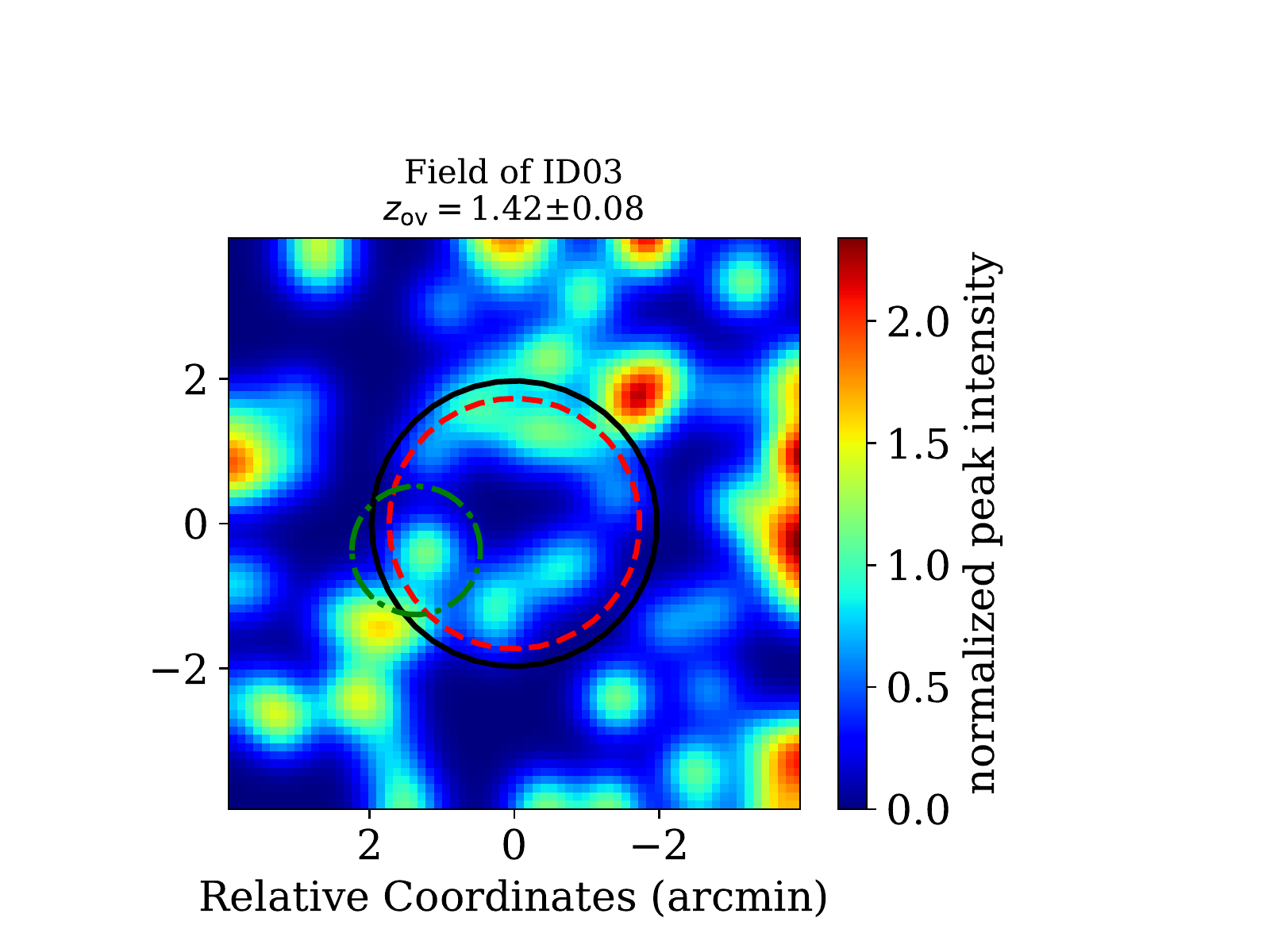}}\\
\vspace{-0.5cm}
\subfloat[]{\hspace{1.cm}\includegraphics[page=16, trim={0.0cm 0.4cm 2cm 1.8cm},clip,width=0.4\textwidth,clip=true]{./ppm_plots_L18.pdf}}
\subfloat[]{\hspace{0.cm}\includegraphics[page=16, trim={0.0cm 0.4cm 0cm 1.0cm},clip,width=0.45\textwidth,clip=true]{./density_maps_Deltaz_0_295_L18_extended_PClist.pdf}}\\
\caption{PPM plots (left) and density maps (right) for the SMGs and corresponding overdensities found using the L18 photo-$z$ catalog. See the text for details about the color coding.}
\label{fig:liu}
\end{figure*}

\begin{figure*}[]\centering
\captionsetup[subfigure]{labelformat=empty}
\ContinuedFloat
\subfloat[]{\hspace{1.cm}\includegraphics[page=13, trim={0.0cm 0.4cm 2cm 1.8cm},clip,width=0.4\textwidth,clip=true]{./ppm_plots_L18.pdf}}
\subfloat[]{\hspace{0.cm}\includegraphics[page=4, trim={0.0cm 0.4cm 0cm 1.0cm},clip,width=0.45\textwidth,clip=true]{./density_maps_Deltaz_0_295_L18.pdf}}\\
\vspace{-0.5cm}
\subfloat[]{\hspace{1.cm}\includegraphics[page=17, trim={0.0cm 0.4cm 2cm 1.8cm},clip,width=0.4\textwidth,clip=true]{./ppm_plots_L18.pdf}}
\subfloat[]{\hspace{0.cm}\includegraphics[page=8, trim={0.0cm 0.4cm 0cm 1.0cm},clip,width=0.45\textwidth,clip=true]{./density_maps_Deltaz_0_295_L18.pdf}}\\
\vspace{-0.5cm}
\subfloat[]{\hspace{1.cm}\includegraphics[page=6, trim={0.0cm 0.4cm 2cm 1.8cm},clip,width=0.4\textwidth,clip=true]{./ppm_plots_L18.pdf}}
\subfloat[]{\hspace{0.cm}\includegraphics[page=1, trim={0.0cm 0.4cm 0cm 1.0cm},clip,width=0.45\textwidth,clip=true]{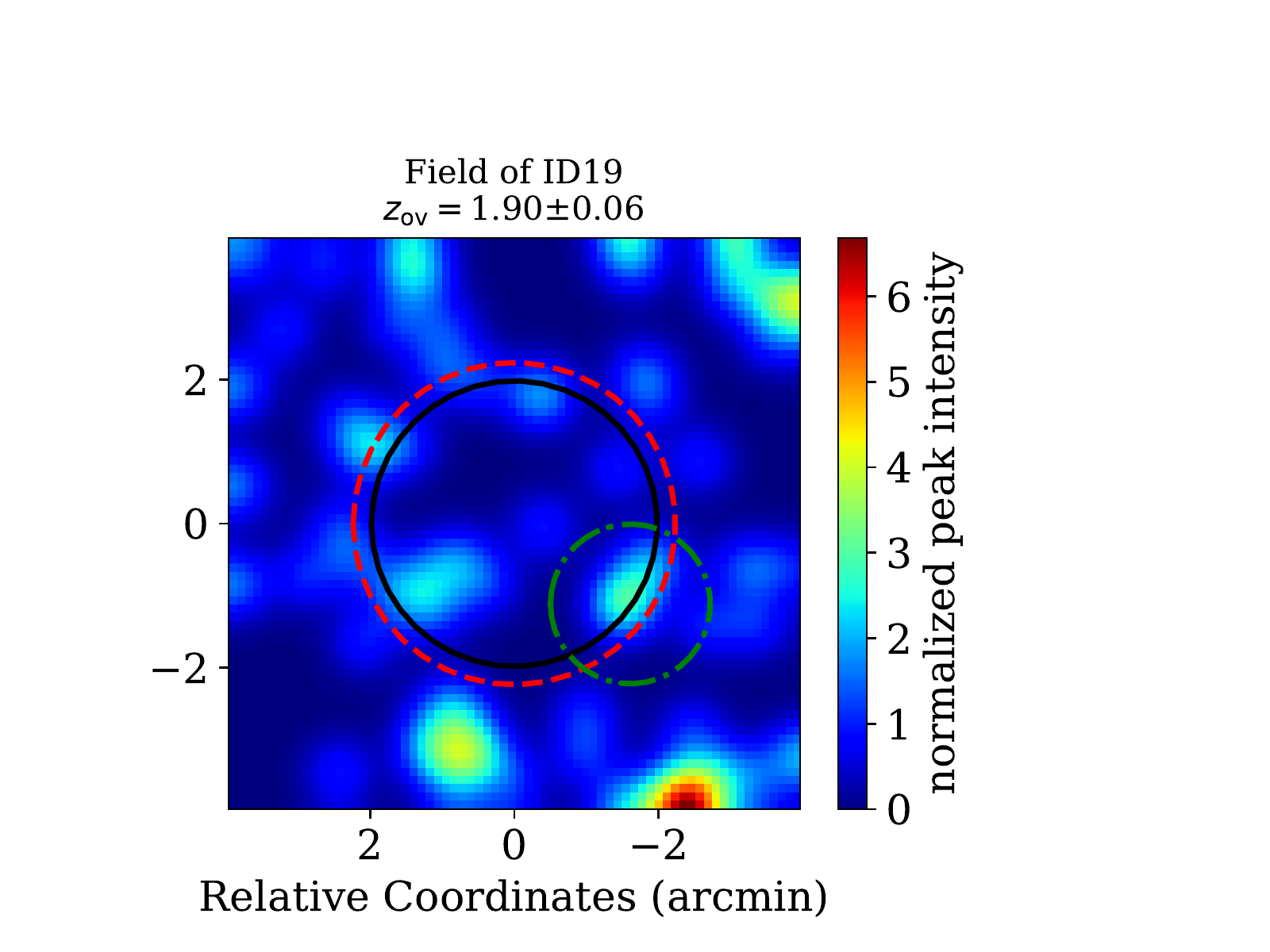}}\\
\vspace{-0.5cm}
\subfloat[]{\hspace{1.cm}\includegraphics[page=15, trim={0.0cm 0.4cm 2cm 1.8cm},clip,width=0.4\textwidth,clip=true]{./ppm_plots_L18.pdf}}
\subfloat[]{\hspace{0.cm}\includegraphics[page=7, trim={0.0cm 0.4cm 0cm 1.0cm},clip,width=0.45\textwidth,clip=true]{./density_maps_Deltaz_0_295_L18.pdf}}\\
\caption{Continued.}
\label{fig:liu}
\end{figure*}

\begin{figure*}[]\centering
\captionsetup[subfigure]{labelformat=empty}
\ContinuedFloat
\subfloat[]{\hspace{1.cm}\includegraphics[page=12, trim={0.0cm 0.4cm 2cm 1.8cm},clip,width=0.4\textwidth,clip=true]{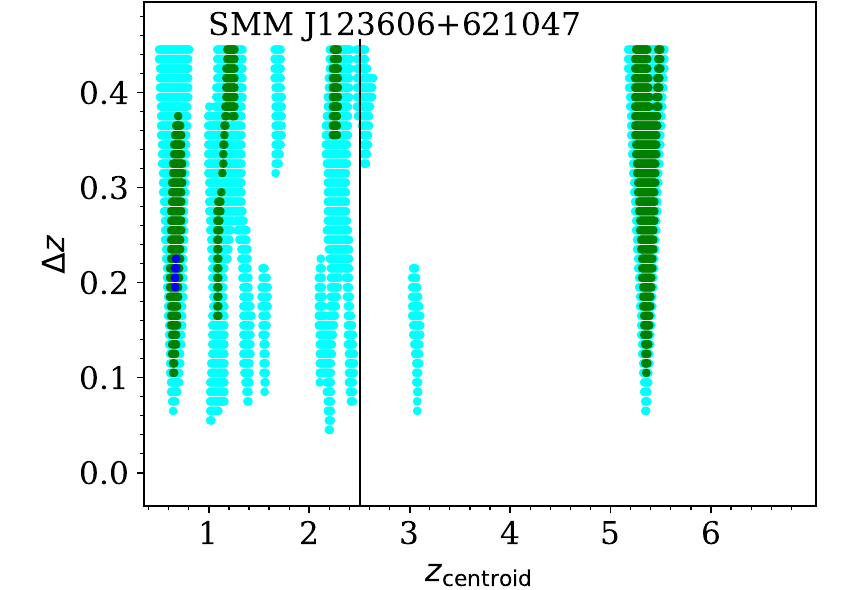}}
\subfloat[]{\hspace{0.cm}\includegraphics[page=3, trim={0.0cm 0.4cm 0cm 1.0cm},clip,width=0.45\textwidth,clip=true]{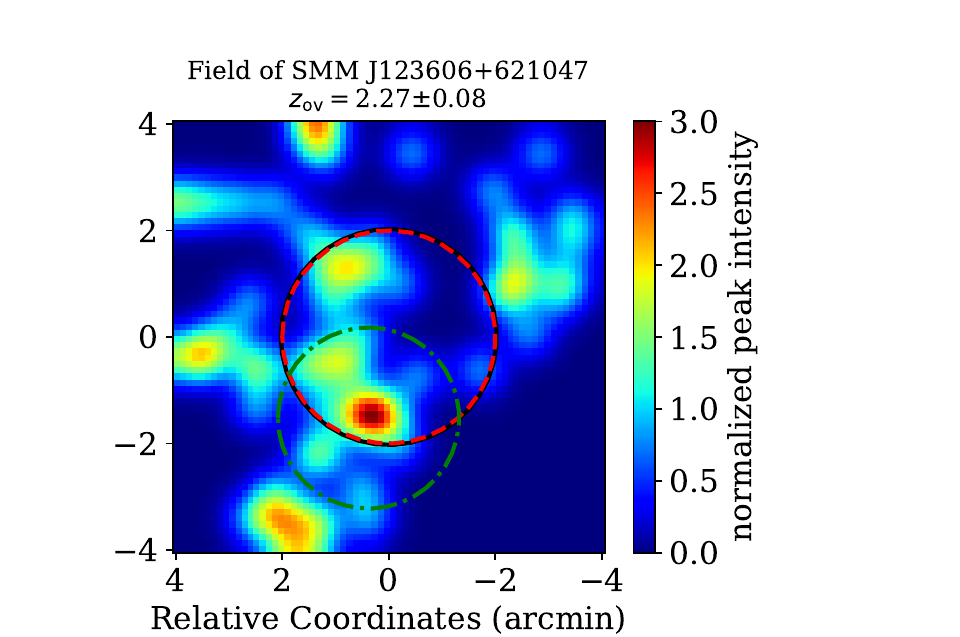}}\\
\vspace{-0.5cm}
\subfloat[]{\hspace{1.cm}\includegraphics[page=8, trim={0.0cm 0.4cm 2cm 1.8cm},clip,width=0.4\textwidth,clip=true]{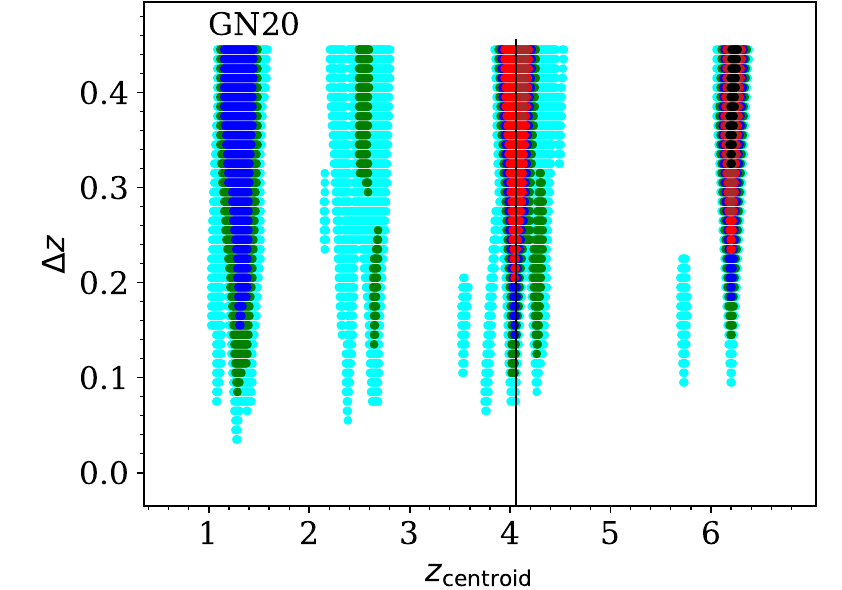}}
\subfloat[]{\hspace{0.cm}\includegraphics[page=1, trim={0.0cm 0.4cm 0cm 1.0cm},clip,width=0.45\textwidth,clip=true]{./density_maps_Deltaz_0_295_L18.pdf}}\\
\vspace{-0.5cm}
\subfloat[]{\hspace{1.cm}\includegraphics[page=9, trim={0.0cm 0.4cm 2cm 1.8cm},clip,width=0.4\textwidth,clip=true]{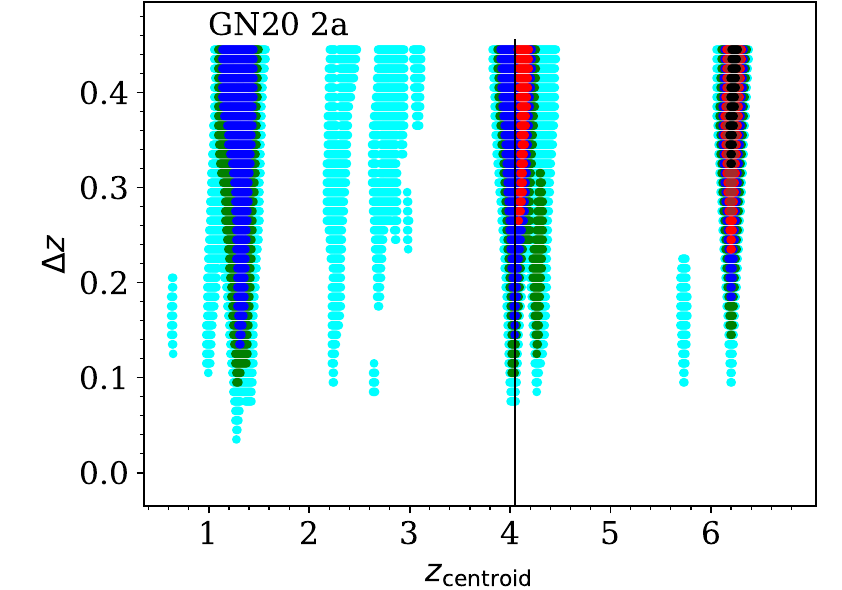}}
\subfloat[]{\hspace{0.cm}\includegraphics[page=2, trim={0.0cm 0.4cm 0cm 1.0cm},clip,width=0.45\textwidth,clip=true]{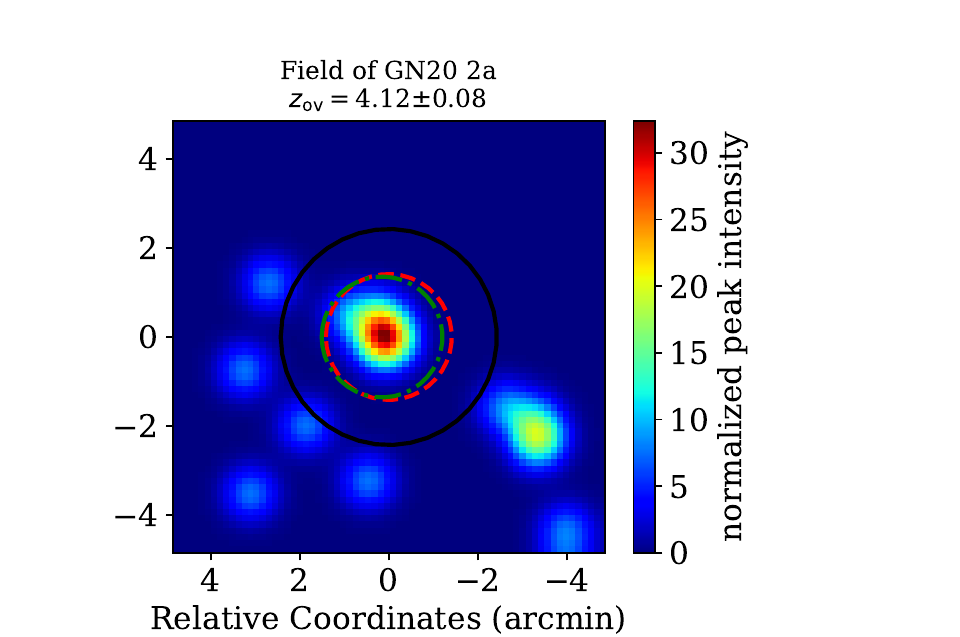}}\\
\caption{Continued.}
\label{fig:liu}
\end{figure*}

\begin{figure*}[h!]\centering
\captionsetup[subfigure]{labelformat=empty}
\subfloat[]{\hspace{1.cm}\includegraphics[page=1, trim={0.0cm 0.4cm 2.0cm 1.8cm},clip,width=0.4\textwidth,clip=true]{./ppm_plots_AH18.pdf}}
\subfloat[]{\hspace{0.cm}\includegraphics[page=1, trim={0cm 0.4cm 0cm 1cm},clip,width=0.45\textwidth,clip=true]{./density_maps_Deltaz_0_295_AH18.pdf}}\\
\vspace{-0.5cm}
\subfloat[]{\hspace{1.cm}\includegraphics[page=7, trim={0.0cm 0.4cm 2.0cm 1.8cm},clip,width=0.4\textwidth,clip=true]{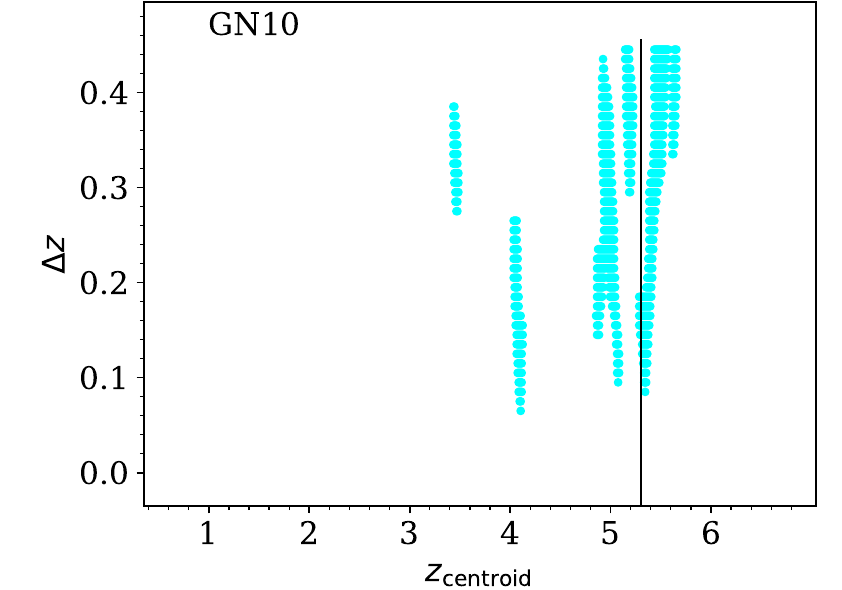}}
\subfloat[]{\hspace{0.cm}\includegraphics[page=2, trim={0.0cm 0.4cm 0.0cm 1.0cm},clip,width=0.45\textwidth,clip=true]{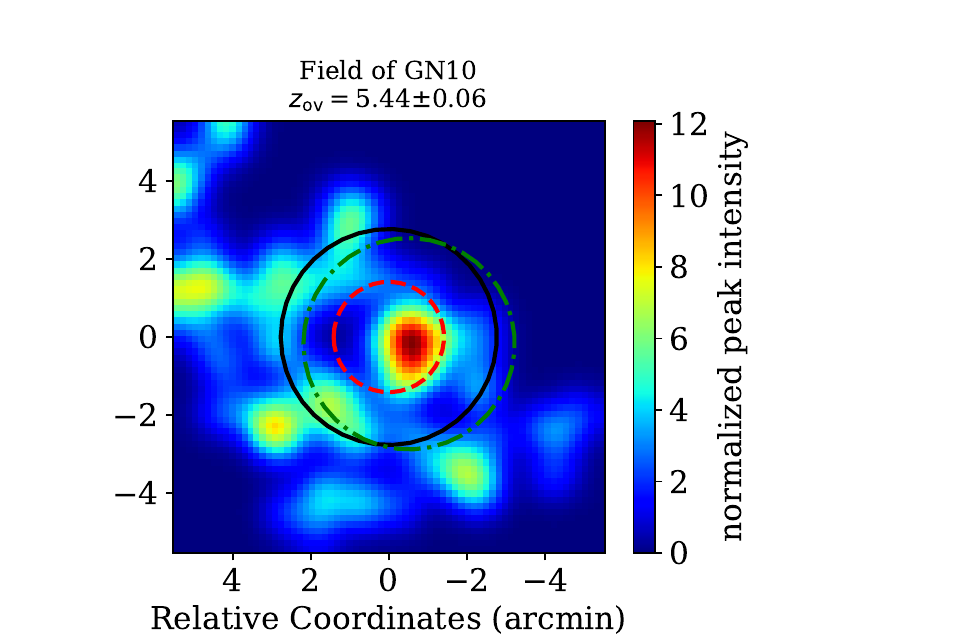}}\\
\caption{PPM plots (left) and density maps (right) for the SMGs and corresponding overdensities found using the AH18 photo-$z$ catalog. See the text for details about the color coding.}
\label{fig:arrabal}
\end{figure*}

\end{appendix}

\end{document}